# Large nonreciprocal absorption and emission of radiation in type-I Weyl semimetals with time reversal symmetry breaking

Yoichiro Tsurimaki[1,†], Xin Qian[1,†], Simo Pajovic [1], Fei Han[2], Mingda Li[2], and Gang Chen[1,*]

[1]Department of Mechanical Engineering, Massachusetts Institute of Technology, Cambridge, MA, 02139, USA

[2]Department of Nuclear Science and Engineering, Massachusetts Institute of Technology, Cambridge, MA, 02139, USA

**Abstract**

The equality between the spectral directional emittance and absorptance of an object under local thermodynamic equilibrium is known as Kirchhoff's law of radiation. The breakdown of Kirchhoff's law of radiation is physically allowed by breaking time reversal symmetry and can open opportunities for nonreciprocal light emitters and absorbers. Large anomalous Hall conductivity and angle recently observed in topological Weyl semimetals, particularly type-I magnetic Weyl semimetals and type-II Weyl semimetals, are expected to create large nonreciprocal electromagnetic wave propagation. In this work, we focus on type-I magnetic Weyl semimetals and show via modeling and simulation that nonreciprocal surface plasmons polaritons can result in pronounced nonreciprocity without an external magnetic field. The modeling in this work begins with a single pair of Weyl nodes, followed by a more realistic model with multiple paired Weyl nodes. Fermi-arc surface states are also taken into account through the surface conductivity. This work points to the promising applicability of topological Weyl semimetals for magneto-optical and energy applications.

[*]Author to whom correspondence should be addressed. E-mail: gchen2@mit.edu.

## I. Introduction

Kirchhoff's law of radiation establishes the equality between the spectral directional absorptance $\alpha_\omega(\mathbf{s})$ and the spectral directional emittance $\epsilon_\omega(-\mathbf{s})$ of an object in local thermodynamic equilibrium, *i.e.*, $\alpha_\omega(\mathbf{s}) = \epsilon_\omega(-\mathbf{s})$, where $\omega$ and $\pm\mathbf{s}$ are the frequency and the direction of incoming and outgoing radiation, respectively. Fundamentally, Kirchhoff's law of radiation underlies the theoretical efficiency limit in radiative energy conversion since converting absorbed incoming radiation into another form of energy, such as electricity or heat, always entails the outgoing emission at the same wavelength in the same direction from the object, which causes an intrinsic loss [1–3]. It has been argued [2,4,5] that Kirchhoff's law of radiation is not a required condition for the validity of the second law of thermodynamics in systems that exchange radiative energy, but rather a result of the Lorentz reciprocity theorem in which the only assumptions are a linear constitutive relation and symmetric permittivity and permeability tensors [6,7]. Thus, the violation of Kirchhoff's law of radiation, *i.e.*, non-reciprocity in the spectral directional absorptance and emittance, is physically allowed, and its realization can open opportunities for light emitters and absorbers for a wide range of radiative applications including solar photovoltaics, thermophotovoltaics, and antennas [1,8].

Non-reciprocity in a medium often arises due to non-zero antisymmetric off-diagonal elements of the dielectric tensor of the medium, which creates non-reciprocal electromagnetic modes [9]. One way to create the antisymmetric off-diagonal elements is by inducing magnetic responses either by the Hall response under an external magnetic field or by spontaneous magnetization in materials, namely the anomalous Hall effect [10]. The anomalous Hall effect can originate from skew scattering and side-jump scattering due to impurities, collectively referred to as extrinsic mechanisms and/or the so-called intrinsic mechanism that is closely related to geometrical properties of the electronic band structure, known as the Berry curvature. In fact, it is known that the anomalous Hall conductivity due to the intrinsic mechanism is given as the integration of the Berry curvature of the occupied Bloch states over the first Brillouin zone [11].

In magneto-optics, the nonreciprocal transmission and reflection of light realized by magneto-optical materials has been widely studied in photonic and plasmonic devices for optical isolators, circulators, and sensing based on magneto-optical Faraday and Kerr effects [9,12–17]. In the context of thermal radiation, the nonreciprocity in directional spectral emittance and absorptance, or equivalently the violation of Kirchhoff's law of radiation, has been predicted by coupling the incident light to non-reciprocal modes created by a magneto-optical material, indium arsenide (InAs), under an external magnetic field of 3 T [2] and 0.3 T [18]. From a practical point of view, however, large nonreciprocity in the emission and absorption spectra without the necessity of external magnetic field is of great interest. However, conventional ferromagnets such as nickel possess a small anomalous Hall angle, and thus the nonreciprocity remains small [19].

Recently, a class of materials called Weyl semimetals has attracted significant attention [20–22]. The linear bulk electronic bands around paired band-crossing points called Weyl nodes near the Fermi energy makes the low-lying excitations behave as relativistic Weyl fermions. Moreover, Weyl semimetals host topologically protected Fermi-arc surface states connecting the projection between the paired Weyl nodes. The topological protection of the Fermi-arc surface states is discussed by considering the projection of the bulk Brillouin zone to the surface Brillouin zone [21]. The corresponding two-dimensional (2D) subsystem is a 2D quantum Hall system and an integer Chern number between Weyl nodes assures the existence of chiral edge modes. In the momentum space, these chiral edge modes form an arc that emerges from a Weyl node of one chirality and terminates into a Weyl node of opposite chirality. The annihilation of Weyl nodes only occurs by meeting these Weyl nodes with different chiral charges in the momentum space, yet their separation is protected as long as either time reversal symmetry (magnetic Weyl semimetals) or spatial inversion symmetry (noncentrosymmetric Weyl semimetals), or both, is broken. Noncentrosymmetric Weyl semimetals have shown potential for optoelectronic applications, *e.g.*, large second-order polarizability for non-linear optics [23], large bulk photovoltaic effect for sensing and energy harvesting [24], and the quantized photogalvanic effect [25]. For nonreciprocal optical responses, however, magnetic Weyl semimetals are of more interest due to nonvanishing integrated Berry curvature of occupied electron states. Moreover, the Weyl nodes are the singular points of the Berry curvature around which the Berry curvature becomes large. This results in a fascinating phenomenon observed in both ferromagnetic and antiferromagnetic Weyl semimetals: the large anomalous Hall effect [26–30]. It is known that conventional ferromagnets such as iron and nickel possess large anomalous Hall conductivity $\sigma_{AHE} \sim 10^3 \; \Omega^{-1} cm^{-1}$ but small anomalous Hall angle $\sigma_{AHE}/\sigma \sim 10^{-3}$, where $\sigma$ is the longitudinal conductivity [31]. In contrast, both large anomalous Hall conductivity and large anomalous Hall angle ranging from 0.01-0.38 were observed in ferromagnetic and antiferromagnetic Weyl semimetals [26–28]. As we will discuss in this work, the energy difference of two surface plasmon modes with opposite propagation direction is proportional to the ratio of off-diagonal to diagonal elements of the dielectric tensor. This ratio can be interpreted as the anomalous Hall angle at finite frequency. Despite the nonmonotonic frequency dependence of the anomalous Hall conductivity, the large anomalous Hall angle is a key to achieve large non-reciprocal optical response. Therefore, Weyl semimetals is a class of materials that is expected to exhibit large non-reciprocal wave propagation [32].

In this work, with an effective model that describes the optical response of both bulk Weyl fermions and surface Fermi-arc states connecting paired Weyl nodes, we show that a class of magnetic Weyl semimetals can be a promising material for nonreciprocal light emitters and absorbers due to its large anomalous Hall effect. Particularly, we show that Kirchhoff's law of radiation can be appreciably broken without an external magnetic field.

**II. Violation of Kirchhoff's law of radiation**

The Kirchhoff's law of radiation is often derived from thermodynamic argument by considering the thermal equilibrium between a black surface and a nonblack surface [33,34]. However, Snyder et al. [5] pointed out the deficiencies in the typical thermodynamic argument considering a nonblack surface in thermodynamic equilibrium with a hemisphere black enclosure, and derived conditions for the Kirchhoff's law of radiation to be valid based on the bidirectional reflectance distribution function of the nonblack surface. Zhu and Fan [2] used a three-surface geometry to reach the same argument when a nonblack surface is specular. Greffet and Nieto-Vesperinas [6] started from coherence theory to generalize the bidirectional reflectance distribution function and discussed the conditions for the validity of the Kirchhoff's law of radiation, arriving at similar results as Snyder et al. [5]. In the following, we will review the conditions to violate the Kirchhoff's law of radiation to arrive at the results of Snyder et al. [5] and Zhu and Fan [2].

We consider a hemispherical enclosure of unit radius that is perfectly reflecting except a small frequency interval $[\omega, \omega + d\omega]$ over which the enclosure is black, and a small nonblack surface area $dA$ at the center of the hemisphere as shown in Fig. 1 (a). We consider the enclosure and the non-black surface are in thermodynamic equilibrium via radiative heat exchange only.

The hemispherical geometry can be used without loss of generality and helps to simplify the derivation. The assumption of thermodynamic equilibrium imposes that net radiative heat flux through any element in the system must be zero.

We consider radiative heat exchange between the nonblack element and a small black element $dA_j$ on the hemisphere as shown in Fig. 1 (b). The radiation that arrives at the element $dA_j$ consists of three components: the emitted radiation from the entire black enclosure (1) directly reaching the element $dA_j$ (excluding the nonblack element and the element $dA_j$ itself); (2) reaching the element $dA_j$ via reflection at the nonblack element $dA$; (3) radiation emitted by $dA$ reaching directly the element $dA_j$. For the first contribution, it can be shown that direct radiative heat exchange between two black elements $dA_i$ and $dA_j$ is net zero due to the geometrical reciprocity of diffuse view factor and hence to all integrated area of $dA_i$. Consequently, the net radiative heat exchange between the black element $dA_j$ and the nonblack element $dA$ must be zero because they are also in equilibrium.

Next, we find the net radiation exchange between $dA$ and $dA_j$ in the presence of the rest of the black enclosure. The thermal radiation reaching $dA_j$ via $dA$ consists of (1) emission from $dA$ and (2) reflected surrounding radiation to $dA_j$ via $dA$ [Fig.1 (b)]. The emission from $dA$ to $dA_j$ is $\epsilon_\omega(\mathbf{s}_j) I_{b\omega}(T) d\omega dA \cos\theta_j \, d\Omega_{dA-dA_j}$ where $\epsilon_\omega(\mathbf{s}_j)$ is the spectral directional emittance of $dA$, $I_{b\omega}(T)$ is blackbody intensity, $\theta_j$ is the polar angle defined in Fig. 1 (a). We introduced a shorthand notation $\mathbf{s}_j = [\sin\theta_j \cos\phi_j, \sin\theta_j \sin\phi_j, \cos\theta_j]^T$ for the direction of the element $dA_j$ in the spherical coordinate. $d\Omega_{dA-dA_j}$ is the solid angle subtended by the element $dA_j$ seen from the element $dA$ and it is equivalent to the area $dA_j$ due to the hemisphere of unit radius, i.e., $d\Omega_{dA-dA_j} = dA_j$. The incident radiation on the nonblack element from the element $dA_i$ is $I_{b\omega}(T)d\omega dA_i d\Omega_{dA_i-dA}$ where $d\Omega_{dA_i-dA} = dA\cos\theta_i$. A fraction of this irradiation is reflected by $dA$ towards $dA_j$: $I_{b\omega}(T)d\omega dA_i d\Omega_{dA_i-dA}\rho_\omega(\mathbf{s}_i \to \mathbf{s}_j)d\Omega_{dA-dA_j}$, where $\rho_\omega(\mathbf{s}_i \to \mathbf{s}_j)$ is the spectral bidirectional reflectance distribution function defined as the ratio of the reflected spectral radiative intensity towards $\mathbf{s}_j$ to the incoming spectral radiative power per unit area normal to the incoming direction $\mathbf{s}_i$. Since radiation emitted from the entire hemisphere can reach to the element $dA_j$ after reflection, we must integrate over the hemispherical solid angle. The sum of the two contributions is the radiosity from the nonblack element and this must be equal to the radiation emitted from the element $dA_j$ towards the nonblack element $I_{b\omega}(T)d\omega dA_j d\Omega_{dA_j-dA}$ where $d\Omega_{dA_j-dA} = dA\cos\theta_j$; which leads to

$$\epsilon_\omega(\mathbf{s}_j) + \int_\Omega \rho_\omega(\mathbf{s}_i \to \mathbf{s}_j)d\Omega_{dA-dA_i} = 1, \tag{1}$$

where $\Omega$ means the entire hemispherical solid angle. Also, we used the reciprocity of diffuse view factor $dA_i d\Omega_{dA_i-dA} = dA\cos\theta_i\, d\Omega_{dA-dA_i}$ to obtain Eq. (1). Note that the solid angle integration over the entire hemisphere includes radiation emitted by $dA_j$ reflected back by the nonblack surface to $dA_j$.

Similarly, the emitted radiation from the black element $dA_j$ towards the nonblack element is partially absorbed and the rest is reflected to the entire hemisphere and we obtain:

$$\alpha_\omega(-\mathbf{s}_j) + \int_\Omega \rho_\omega(\mathbf{s}_j \to \mathbf{s}_i)d\Omega_{dA-dA_i} = 1, \tag{2}$$

where $\alpha_\omega(-\mathbf{s}_j)$ is the spectral directional absorptance of the nonblack element. From Eqs. (1) and (2), we obtain:

$$\epsilon_\omega(\mathbf{s}_j) - \alpha_\omega(-\mathbf{s}_j) = \int_\Omega [\rho_\omega(\mathbf{s}_j \to \mathbf{s}_i) - \rho_\omega(\mathbf{s}_i \to \mathbf{s}_j)]d\Omega_{dA-dA_i}. \tag{3}$$

This is the result Snyder et al. [5] obtained. When we assume that the surface of the nonblack element is specular, then the elements $dA_i$ and $dA_j$ must be on the same plane of incidence ($\phi_j = \phi_i + \pi$) and also the reflection angle must be symmetric ($\theta_i = \theta_j$). Therefore, the radiative heat exchange occurs between only the three elements. Then, Eq. (3) becomes:

$$\epsilon_\omega(\mathbf{s}_j) - \alpha_\omega(-\mathbf{s}_j) = [\rho_\omega(\mathbf{s}_j \to \mathbf{s}_i) - \rho_\omega(\mathbf{s}_i \to \mathbf{s}_j)]d\Omega_{dA-dA_i} = r_\omega(\mathbf{s}_j \to \mathbf{s}_i) - r_\omega(\mathbf{s}_i \to \mathbf{s}_j). \quad (4)$$

where $r_\omega = \rho_\omega d\Omega$ is the spectral directional reflectance of the nonblack surface $dA$. This is the result Zhu and Fan obtained [2]. In the system obeying the Lorentz reciprocity, the bidirectional reflectance distribution function is reciprocal $\rho_\omega(\mathbf{s}_j \to \mathbf{s}_i) = \rho_\omega(\mathbf{s}_i \to \mathbf{s}_j)$ and the Kirchhoff's law of radiation holds $\epsilon_\omega(\mathbf{s}_j) = \alpha_\omega(-\mathbf{s}_j)$ [4,6,35]. However, if one realizes a nonreciprocal reflector $\rho_\omega(\mathbf{s}_j \to \mathbf{s}_i) \neq \rho_\omega(\mathbf{s}_i \to \mathbf{s}_j)$ by breaking the Lorentz reciprocity, the violation of the Kirchhoff's law of radiation results without breaking the second law of thermodynamics. Magnetic Weyl semimetals provide a platform to achieve this by breaking time reversal symmetry.

Previous thermodynamic arguments missed [33,34] the fact that the total radiative heat exchange between $dA_j$ and $dA$ must be zero when both direct and reflected radiative heat fluxes are included, but individual contributions from reflection, emission, and absorption can be different. Rather, radiation exchange between $dA_j$ and $dA$ was assumed equal [33] or the reciprocity in reflection was introduced without further discussion [34], both of which lead to the Kirchhoff's law of radiation as is conventionally understood.

### III. Dielectric function model

Several models of the dielectric tensor of type-I Weyl semimetals have been proposed. The approach in [32,36] is based on the constitutive relation of the electric displacement field that includes two additional terms describing the anomalous Hall current and the chiral magnetic effect, both of which contribute to off-diagonal elements of the dielectric tensor [36,37]. Both effects are the manifestations of the chiral anomaly, *i.e.*, nonconservation of the chiral current, and are described by the axion term in the electromagnetic field action [38]. The Weyl node separation in the momentum space is effectively a magnetic field in momentum space acting on fermions and induces the off-diagonal element $\sigma_{AHE}(\omega)/\omega = 2ie^2|\mathbf{b}|/\pi\hbar\omega$, where $\pm\mathbf{b}$ is the locations of the Weyl nodes in the momentum space. Although this expression is derived for the Fermi energy located at the Weyl node, this intrinsic contribution dominates over the extrinsic contribution as long as the Fermi energy is small enough so two Fermi surfaces of two Weyl nodes do not merge [39]. For the diagonal terms, the one-band model that only describes the intra-band contribution [36] as well as the two-band model that includes the inter-band transition [32] was proposed. Recently, the dielectric tensor of noncentrosymmetric Weyl semimetals TaAs and NbAs, was also calculated from a first-principles approach based on density functional theory [40] and showed good agreements with experiments [41,42]. The dielectric tensor is also studied with the Kubo formula for effective Weyl Hamiltonians [43,44].

In this work, our modeling of the local dielectric tensor of a type-I magnetic Weyl semimetal is based on the work reported in [43]. Our effective Hamiltonian describes two connected Weyl cones [45,46] $H(\mathbf{k}) = \hbar v_F\big((k_x^2 + k_y^2 - m^2)/2b\hat{\sigma}_x + k_y\hat{\sigma}_y + k_z\hat{\sigma}_z\big)$, where $v_F$ is the Fermi velocity, $\hat{\sigma}_i$ ($i = x, y, z$) are the Pauli matrices and $m$ is the control parameter of the effective Hamiltonian, and we let $m = b^2$. The two Weyl nodes are separated in the $k_x$-direction by $2b$. The quadratic terms are introduced in order to connect the two Weyl cones. The electronic band dispersion on the surface $k_z = 0$ of the effective Hamiltonian is shown in Fig. 2 (a). Although the effective Hamiltonian that contains quadratic terms in all directions is more general, we only consider the quadratic terms in $k_x$ and $k_y$ as they already capture the essential physics. The eigenstates of the bulk Weyl fermions and the Fermi-arc surfaces states described by the effective Hamiltonian can be analytically derived. Once the eigenstates are known, the optical conductivity is calculated by the Kubo formula. The model gives both the bulk and surface optical conductivities. The bulk optical conductivity includes both intra-band and inter-band transitions of bulk Weyl fermions and the surface conductivity includes the surface-to-surface and surface-to-bulk transitions due to the Fermi-arc surface states. The bulk dielectric tensor from a single pair of Weyl nodes when the Weyl node separation is in $k_x$-direction is written as:

$$\hat{\varepsilon}(\omega) = \begin{bmatrix} \varepsilon_{xx}(\omega) & 0 & 0 \\ 0 & \varepsilon_{yy}(\omega) & ig \\ 0 & -ig & \varepsilon_{zz}(\omega) \end{bmatrix}, \quad (5)$$

where the diagonal terms are related to the bulk optical conductivity $\sigma(\omega)$ via $\varepsilon_{nn}(\omega) = \varepsilon_b + \frac{i\sigma_{nn}(\omega)}{\varepsilon_0 \omega}$ ($n = x, y, z$) and $g = \sigma_{yz}(\omega)/\varepsilon_0 \omega$ is the contribution from the anomalous Hall effect. $\varepsilon_b$ is the background dielectric constant that accounts for contributions from free carriers in other bands as well as dielectric response at high frequencies. $\varepsilon_0$ is the vacuum permittivity. We take $\varepsilon_b=10$, but our results have little to do with this choice. In calculations of the dielectric tensor, we set the temperature to $T=0$ K throughout our calculation to obtain a closed-form expression for dielectric tensor [43]; The incorporation of the temperature dependence is straightforward.

Weyl semimetals possess topologically-protected Fermi-arc surface states. In our model, the eigenstates of the Fermi-arc states can be analytically calculated. Using those eigenstates, the optical surface conductivity tensor $\hat{\sigma}^S(\omega)$ is calculated by the Kubo formula by including the surface-to-surface and surface-to-bulk transitions. The surface conductivity tensor $\hat{\sigma}^S(\omega)$ has a similar structure to the bulk dielectric tensor, i.e., non-zero $\sigma_{yz}^S(\omega)$. In our optical simulation, we model the presence of the Fermi-arc states as the existence of surface charge and surface current, which is directly incorporated in the electromagnetic interface conditions. For the anomalous Hall effect term $g$, the effective Hamiltonian only accounts for the contribution from the vicinity of the Weyl nodes, which may be considered as an underestimation of the anomalous Hall conductivity since the larger contribution to the anomalous Hall conductivity can originate from nodal lines far from the Weyl nodes, as observed in some ferromagnetic Weyl semimetals such as $Co_2Sn_2S_2$ [27]. The expressions of the bulk dielectric tensors and the surface conductivity tensors used in this work can be found in the Supplemental Material, Sec. 1 [47].

In the dielectric tensor model, five parameters ($\mathbf{b}$, $v_F$, $E_F$, and $\gamma_b$, and $\gamma_s$) need to be determined. Here $\gamma_b$ and $\gamma_s$ are dissipative losses of bulk Weyl fermions and Fermi-arc fermions due to interactions. To reasonably select the parameters, we take relevant values from the literature. We consider two Fermi energies above the Weyl nodes: $E_F = 60$ meV and $E_F = 100$ meV. Those magnitudes of the Fermi energy are comparable to those of magnetic Weyl semimetals including $Co_3Sn_2S_2$ ($E_F\sim 60$meV) [26,27], $Co_2MnGa$ ($E_F\sim 80$meV) [48]. Existing literature predicted the existence of many other magnetic Weyl semimetals [49–51]. We set the Fermi velocity to be $v_F = 1.0 \times 10^5$ m/s, which is also in the range of theoretically and experimentally determined values in ferromagnetic Weyl semimetals including $Co_3Sn_2S_2$ [26,27] ($v_F\sim 1.3 \times 10^5$m/s), $Co_2MnGa$ ($v_F\sim 1.2 \times 10^4$m/s) [48], $Y_2Ir_2O_7$ ($2 \times 10^5$m/s)[41], and $Eu_2Ir_2O_7$ ($4 \times 10^5$m/s) [53]. We choose the Weyl node separation $2b = 0.45$Å$^{-1}$, also close to the separations observed in the above materials. With these parameters, the electronic band structure on the $k_x$ axis ($k_y = k_z = 0$) is shown in Fig. 2 (b) for the two Fermi energies. At $E_F = 60$ meV, the Fermi surface consists of two separate surfaces, each of which enclosing a single Weyl node. In this case, the intrinsic mechanism of anomalous Hall conductivity (DC limit) has been shown to dominate [39] and $\sigma_{AHE} = 2e^2|\mathbf{b}|/\pi\hbar$ is good approximation. With our Fermi velocity, the two Fermi surfaces from the two Weyl nodes merge above ~80 meV. Thus, at Fermi energy $E_F = 100$ meV, the Fermi surface is one sheet across the two Weyl nodes and the anomalous Hall conductivity decreases drastically at low frequencies (Supplemental Materail, Sec. 2 [47]). The purpose of selecting $E_F = 100$ meV is to study the nonreciprocal behavior above this point where the anomalous Hall conductivity becomes smaller. The dissipative loss of bulk Weyl fermions $\gamma_b$ and surface Fermi arc states $\gamma_s$, which is formally given as the imaginary part of the electron self-energy, is less straightforward to obtain. In our model of the dielectric tensor, we set $\gamma = \gamma_b = \gamma_s = 1.5$ meV and $\gamma = \gamma_b = \gamma_s = 4.2$ meV for $E_F = 60$ meV and $E_F = 100$ meV, respectively. Depending on interactions and disorders that exist in the Weyl semimetal, the magnitude of the dissipative loss can vary. Taking these representative parameters, we study the optical responses of magnetic Weyl semimetals.

Figure 2 (c-f) show the real and imaginary parts of the bulk dielectric tensor and surface conductivity tensor elements from a pair of Weyl nodes in the medium. The Fermi energy is 60 meV. Those for $E_F = 100$ meV are shown in Supplemental Material, Sec. 3 [47]. The real parts of the dielectric tensor elements have peaks at the onset of the inter-band transition, $E = 2E_F$, and the imaginary parts become large around this onset frequency due to the absorption of light. At low frequencies, the contributions from the intra-band transition dominate, and the dielectric tensors show Drude-like responses. The surface conductivity elements are normalized by $e^2/h$. As we will show later, the existence of Fermi-arc surface states can create additional channels for light absorption and can alter the dispersion relation of surface plasmons due to bulk Weyl fermions. The dispersion relation of Fermi-arc surface plasmons and their coupling to bulk plasmons have been studied previously [54–57].

## IV. Optical grating structure

In section II, we discussed that the violation of Kirchhoff's law of radiation results from the nonreciprocity of spectral directional reflectance when the surface is specular. In a translationally invariant and optically thick system, no fraction of incident light transmits through the system, and the nonreciprocity in reflection is also equivalent to the nonreciprocity in spectral directional absorptance from two axisymmetric direction $\alpha_\omega(\theta) \neq \alpha_\omega(-\theta)$. Furthermore, in order to assure that the incoming radiation from the direction $\theta$ is only reflected into the direction $-\theta$, we design a structure that diffracts only zeroth-order mode. Under these conditions, we show the nonreciprocity in spectral directional absorptance from two axisymmetric angles to discuss the breakdown of Kirchhoff's law of radiation. In this work, we design a structure that supports nonreciprocal surface plasmon polaritons (SPPs) at the interface of a dielectric and a magnetic Weyl semimetal. We consider a structure made of a low-loss dielectric grating with a nondispersive and isotropic dielectric constant $\varepsilon_d = 10 + i0.01$ on top of a magnetic type-I Weyl semimetal as shown in Fig. 3. The entire structure is in air. It is known that a non-reciprocal SPP mode exists in the Voigt configuration, where the spontaneous magnetization or external magnetic field in the *x*-direction is perpendicular to the direction of SPP propagation in the *y*-direction [36,58]. Also, the direction of the spontaneous magnetization is almost parallel to the Weyl node separation direction. Thus, we consider the crystal orientation of the magnetic Weyl semimetal such that its spontaneous magnetization is in parallel to the surface and directed to the *x*-direction. We focus on *p*-polarized light (magnetic field pointing the *x*-direction) as only it excites SPPs. Inside the bulk of Weyl semimetal, however, all components of the electromagnetic fields can be non-zero. The dispersion relation of SPPs can be derived by seeking an exponentially decaying electric field in both air and the Weyl semimetal such that the electric field satisfies the electromagnetic interface conditions. By extending the dispersion relation of the SPPs with a uniaxial dielectric tensor as discussed in [58] to the biaxial dielectric tensor, we obtain the dispersion relation of SPPs as:

$$\varepsilon_d(q^2 - \varepsilon_{zz}k_0^2) + gq\gamma_0 + \varepsilon_{zz}\gamma_1\gamma_0 = 0, \tag{6}$$

where $\varepsilon_d$ is the dielectric constant of the dielectric, $\gamma_0^2 = q^2 - \varepsilon_d k_0^2$ and $\gamma_1^2 = \varepsilon_{zz}/\varepsilon_{yy}q^2 - (\varepsilon_{yy}\varepsilon_{zz} - g^2)k_0^2/\varepsilon_{zz}$ are the perpendicular components of the wavevectors in the dielectric and the Weyl semimetal, respectively. $k_0$ is the wavevector in vacuum. The dispersion relation contains a term proportional to $q$, which creates the nonreciprocal propagation $\omega(q) \neq \omega(-q)$. Clearly, in the absence of the anomalous Hall effect ($g = 0$) the dispersion is quadratic in $q$, replenishing the reciprocity $\omega(q) = \omega(-q)$. In the limit $|q| \gg k_0$, the dispersion relation Eq. (6) can be approximated as $|\varepsilon_{yy}(\omega)| \pm g - \varepsilon_d = 0$. Therefore, the energy difference of the two counter-propagating SPPs is wider as the ratio $\text{Re}[g]/|\varepsilon_{yy}(\omega)|$ becomes large. At low frequency range where $\varepsilon_b$ is small compared to the intra-band contribution, then the ratio above is interpreted as the anomalous Hall angle at finite frequency. Since the largely separated frequencies of the two non-reciprocal SPPs in the limit $|q| \gg k_0$ will give larger frequency window in which the two SPP branches have well-separated propagation constants, materials with a large $\text{Re}[g]/|\varepsilon_{yy}(\omega)|$ are suitable to achieve near complete violation of Kirchhoff's law of radiation without an external magnetic field. Although we discussed the behavior in the limit $|q| \gg k_0$, the propagation wavevector is still much smaller than the Fermi wavevector $|q| \ll k_F$, thus the nonlocality of the dielectric function is negligible.

Besides the SPPs and the Fermi-arc surface states, the bulk plasmons are supported in the Weyl semimetal. The dispersion relations of the bulk plasmons are obtained from the wave propagation equation $\boldsymbol{n}(\boldsymbol{n} \cdot \boldsymbol{E}) - n^2 \boldsymbol{E} - \hat{\varepsilon}(\omega)\boldsymbol{E} = 0$ where $\boldsymbol{n} = c\boldsymbol{k}/\omega$ and $\boldsymbol{E}$ is the electric field in the Weyl semimetal. From the wave propagation equation, the dispersion relation of bulk plasmons propagating in the y-direction ($n_x = n_z = 0, n_y \neq 0$) in the Weyl semimetal are given by:

$$\varepsilon_O = \varepsilon_{xx}, \quad \varepsilon_X = (\varepsilon_{yy}\varepsilon_{zz} - g^2)/(\cos^2\phi\, \varepsilon_{zz} + \sin^2\phi\, \varepsilon_{yy}) \tag{7}$$

where $\varepsilon_O$ and $\varepsilon_X$ are the ordinary and extraordinary modes, respectively and $\phi$ is the polar angle of propagation inside the Weyl semimetal.

## V. Results and Discussion

### A. Single Weyl pair without the Fermi-arc states

First, we consider the case of a prototypical Weyl semimetal that possesses a single pair of Weyl nodes. Figure 4 (a) shows the dispersion relations of the SPPs at the interface of the dielectric and such a Weyl semimetal, as well as those of the bulk plasmons (blue-shaded region) with respect to the effective refractive mode index $n_{eff}$. The effective refractive index is defined as $q = n_{eff}k_0$, where $q$ is the wavevector along the propagation direction. The Fermi energy is 60 meV. Note that this dispersion relation does not include the Fermi-arc surface state and the dispersion relation of SPPs is calculated assuming no dissipative loss. At low frequencies, the effective refractive mode index is close to the light line of the dielectric ($\mathrm{Re}[n_{eff}] \sim \sqrt{\mathrm{Re}[\varepsilon_d]} \sim 3.16$). At frequencies above $E/E_F \sim 0.35$, or equivalently ~5 THz, the two SPP modes start to show non-reciprocity. The SPPs propagating in the negative y-direction continue to be supported at higher energy until they merge with the extraordinary bulk plasmon mode (leaky mode). The green lines are the dispersion of the grating equation $n_{eff} = \sin\theta + m\lambda/\Lambda$ at the incident angles of $\theta = \pm 60°$ and the diffracting orders of $m = \pm 1$ for the case of $\Lambda = 20\ \mu m$, respectively. As seen, we expect that the SPPs excitation occurs at around $E/E_F \sim 0.37$ and 0.41 for $\theta = +60°$ and $-60°$, respectively. In reality, the frequencies will shift due to the imaginary parts of the bulk dielectric tensor elements. With the model dielectric functions, we solved the frequency-domain Maxwell's equations in the grating structure by the finite-element method to numerically obtain the spectral directional absorptance of the grating structure. We first consider the case without the presence of the Fermi-arc surface state. By tuning the height and thickness of the grating $h$ and $t$, we can design the structure to achieve complete absorption at the incident angle of $\theta = -60°$ as shown in Fig. 4 (b). The spectral directional absorptance for the angle of incidence $\theta = 60°$ still shows high absorptance of $\sim 0.8$ at a lower frequency by $\Delta E \sim 0.03 E_F$ due to the non-reciprocal SPPs. The geometrical parameters of the structure are stated in the caption of Fig. 4. As shown, the Weyl semimetal exhibits large non-reciprocity without an external magnetic field. Moreover, since the structure is optically-thick and only supports the zeroth-order diffraction, the difference in the spectral directional absorptance is equivalent to the difference in the spectral directional reflectance, thereby violating Kirchhoff's law of radiation [2]. One may expect even larger nonreciprocity at frequencies above $E/E_F \sim 0.4$, as the dispersion relation predicts unidirectional propagation of SPPs. However, the Weyl semimetal becomes absorptive due to the bulk plasmons at those frequencies as shown in Fig. 4 (a), and the absorption peaks from nonreciprocal SPP modes are inundated in the high bulk absorptance. In fact, the spectral directional absorptance of the grating structure at higher frequencies in Fig. 4 (b) shows similar trend to the spectral directional absorptance of bulk Weyl semimetal shown in Fig. 4 (a) since bulk plasmons absorption is dominant. Thus, we must design the structure at the frequencies where the nonreciprocal propagation is evident, yet the Weyl semimetal is not lossy. Similarly, the nonreciprocal absorption spectra can be achieved at the Fermi energy of 100 meV as shown in Fig. 4 (c) and (d) with the larger peak separation $\Delta E \sim 0.04 E_F$. For both cases, the large nonreciprocity without external magnetic field is due to the occurrence of the coupling to the SPP modes at the frequency range where the ratio remains large: $\text{Re}[g]/\text{Re}[\varepsilon_{yy}] \sim 0.6$. At lower frequencies, the Weyl semimetal becomes more reflective, which is favorable for achieving resonant absorption with high quality factor, but the non-reciprocity becomes small as $\text{Re}[g]/\text{Re}[\varepsilon_{yy}]$ is small. On the other hand, at higher frequencies, the non-reciprocity becomes large, but the Weyl semimetal becomes lossy, thereby the two absorption peaks overlap over a wide range of frequencies. As a consequence, one needs to find the optimized condition at the frequency region in between the two.

### B. Single Weyl pair with the Fermi-arc states

Next, we include the Fermi-arc surface states in our simulation. The Fermi arc surface states exist for the parallel wavevector component $k_x^2 + k_y^2 \leq b^2$. Since the propagation constant of nonreciprocal SPPs is $q = n_{eff} k_0 \sim 10^5 - 10^6 \text{ m}^{-1}$, the Fermi arc surface states can be simultaneously supported by the structure and will alter the behavior of SPPs. In our optical simulation, we incorporate the effects of the Fermi-arc states via the interface conditions of the electromagnetic field (Supplemental Material, Sec. 4 [47]). The finite surface conductivity tensor due to the existence of the Fermi-arc surface states creates surface charge as well as surface current components parallel to the surface, thereby making the tangential components of magnetic fields no longer continuous. Specifically, the magnetic field components $H_x$ and $H_y$ jump by the amount of the surface current components parallel to the surface $i_{e,x}^S = \sum_i \sigma_{xi}^S E_i$ and $i_{e,y}^S = \sum_i \sigma_{yi}^S E_i$. Moreover, the surface current component normal to the surface $i_{e,z}^S = \sum_i \sigma_{zi}^S E_i$ is finite due to non-zero $\sigma_{zy}^S$ and $\sigma_{zz}^S$, which creates dipole moments normal to the surface. As a result, the tangential components of electric field $E_x$ and $E_y$ also jump by the spatial derivative of the normal surface current $i/\varepsilon_0 \omega \partial_x i_{e,z}^S$ and $i/\varepsilon_0 \omega \partial_y i_{e,z}^S$. Figure 5 (a) and (b) show the spectral directional absorptance of the grating structure at the Fermi energies 60 meV and 100 meV in the presence of the Fermi-arc surface states. The presence of the Fermi-arc surface states creates additional absorption channels and broadens the spectral directional absorptance peaks. Also, the absorption peak slightly red-shifts for $E_F = 60$ meV while it does not affect the peak position when $E_F = 100$ meV. The shift of absorption peak is more prominent at lower Fermi energy.

### C. Multiple Weyl pairs

Finally, we consider a more realistic case in which a Weyl semimetal possesses multiple Weyl nodes. Previously, multiple paired Weyl nodes were modeled by simply multiplying the diagonal components of the optical conductivity tensor of a single Weyl cone by the number of Weyl cones, while keeping the anomalous Hall conductivity the same as the contribution from single Weyl pair [32]. In realistic Weyl semimetals, multiple pairs of Weyl nodes are not oriented in the same direction. In this work, we include relative orientations of Weyl node pairs in the model of bulk and surface conductivity tensors to investigate their effects on the dielectric response and the spectral absorptance. As a representative material for taking account for relative orientation of Weyl node pairs, we use the Weyl node locations and its relative orientations of half-metallic Heusler ferromagnet $Co_3Sn_2S_2$. There are a total six Weyl nodes in this material and it is numerically found that a Weyl node is located at (0.360922, −0.059795, −0.059809) [27] in terms of fractional coordinates of the rhombohedral reciprocal structure. The locations of the other five Weyl nodes can be determined by $C_3$ rotational symmetry and inversion symmetry as shown in the inset of Fig. 6 (a). By transforming to the orthonormal coordinates where one of the pair of Weyl nodes is aligned with $k_x$-axis, we can represent the other two Weyl nodes in the orthonormal coordinate via rotation (Supplemental Materaisl, Sec. 5 [47]). As a result, we model the total bulk dielectric tensor as the summation of the contributions from the three oriented Weyl pairs: $\hat{\varepsilon}(\omega) = \varepsilon_b \hat{I}_3 + \hat{\varepsilon}_{Weyl}(\omega) + R_2 \hat{\varepsilon}_{Weyl}(\omega)(R_2)^T + R_3 \hat{\varepsilon}_{Weyl}(\omega)(R_3)^T$, where $\hat{I}_3$ is the identity matrix and $\hat{\varepsilon}_{Weyl}$ is the dielectric tensor of a single Weyl node pair given in Eq. (5) and $R_2$ and $R_3$ are the matrices that transform the other two pairs of Weyl nodes. Note that all the elements of the total dielectric tensor in this model are non-zero, thereby we must include all electric and magnetic field components in the interface conditions. Figure 6 (a) and (b) show the comparison of the diagonal (*yy*) and off-diagonal (*yz*) components of the dielectric tensor in our model and those elements obtained by the simple multiplication, *i.e.*, $\hat{\varepsilon}(\omega) = \varepsilon_b \hat{I}_3 + 3\hat{\varepsilon}_{Weyl}(\omega)$. It can be seen that the diagonal elements are not affected by the different orientation of Weyl node pairs while the off-diagonal elements can result in an overestimation if one uses the simple multiplication. The surface conductivity tensor is also modeled in a similar manner. All tensor elements in our model are shown in Sec. 6 of the Supplemental Material [47]. Figure 6 (c) and (d) show the absorption spectra for $E_F$=60 meV and $E_F$=100 meV, respectively. Both figures show the comparison between the cases with and without the presence of the Fermi arc surface states. The structure can be designed so it exhibits resonant absorption peaks, and even in this realistic model, the nonreciprocity is still appreciable. Therefore, we expect that the nonreciprocity can be experimentally observed in realistic magnetic Weyl semimetals. Moreover, our model takes into account the contributions to the anomalous Hall conductivity only from the vicinity of the Weyl node pairs. The inclusion of the contributions from the entire Brillouin zone can lead to more prominent nonreciprocity. In this work, we assumed the zero temperature for the modeling of the dielectric and surface conductivity tensors, but we note that there could exist quantitative differences at higher temperatures. First, the Fermi-Dirac distribution is no longer a step function, which will give small corrections to the bulk dielectric and surface conductivity tensors. Second, at higher temperatures, the longitudinal conductivity decreases due to stronger scattering, while the anomalous Hall conductivity remains almost the same. This will lead to a larger Hall angle as observed in the experiment [26], which may favor large nonreciprocity. However, the dissipative loss also becomes larger at higher temperatures. Thus, for a working temperature of interest, the design of the grating parameters accounting for the two competing effects will be required. In the modeling of the bulk dielectric surface conductivity tensors, we used two representative values for the fermion lifetime $\tau = 1/\gamma$. Detailed effects of interactions and disorder can be included by calculating the fermion self-energy under interactions of interest such as quenched impurities [39], Coulomb interactions [59], and inter-Weyl-nodes scattering [60,61]. As a lifetime of Weyl fermions becomes shorter, the dissipative loss $\gamma$ becomes larger. As a result, the resonant absorption peaks observed in our results become broader. Finally, although the transition to magnetic Weyl semimetal phases occurs at low temperature in some ferromagnetic Weyl semimetals, such as $Co_3Sn_2S_2$ ($T = 170$ K) [26,62] and $Eu_2I_2O_7$ ($T = 110$ K), and may not be suitable for high temperature applications, other ferro- and ferrimagnetic Weyl semimetals such as $Ti_2MnAl$ [63] and $Co_2MnGa$ [48,64] possess Curie temperatures over 650K, which would be more suitable for applications above room temperature.

## VI. Conclusions

In summary, our modeling shows that a class of type-I magnetic Weyl semimetals can exhibit large nonreciprocal emission and absorption spectra of radiation, enabling the breakdown Kirchhoff's law of radiation without applying external magnetic field. We illustrate that large nonreciprocity and high background absorptance are competing effects in the design of non-reciprocal resonant light emitters and absorbers. We also discuss the presence of Fermi-arc surface states that broaden the absorption peak and slightly red-shifts the SPP dispersion at the lower Fermi energy $E_F = 60$ meV, but the non-reciprocal absorption spectra are not affected much in the frequency range of our interest. The relative orientation of multiple Weyl node pairs is modeled, and we show that the off-diagonal elements of the dielectric tensor can be affected. Since this can change SPP dispersion, the consideration will be required when one designs a structure to achieve the absorption at the desired frequency. Finally, our study points out the possibility of using magnetic Weyl semimetals for nonreciprocal light emitters and absorbers.

**Acknowledgement**: We acknowledge Dr. Bo Zhao, Mr. Cheng Guo, and Professor Shanhui Fan with whom we had good exchange on similar ideas (see ArXiv.1911.08035 and our arXiv paper). Professor Fan's paper was subsequently published in Nano Letter (Nano Lett. 2020, 20, 3, 1923-1927). Our submission to Nano Letter was not sent out for review. This work is supported by ARO MURI (Grant No. W911NF-19-1-0279) via U. Michigan.

† Y.T. and X.Q. contributed equally to this work.

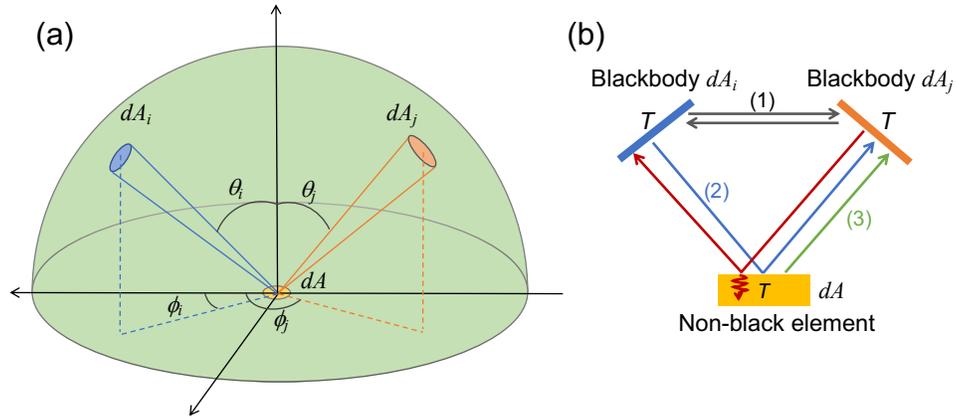

Figure 1: (a): A black hemispherical enclosure and a small nonblack surface in thermodynamic equilibrium. Positions of two surface elements $dA_i$ and $dA_j$ are shown in the spherical coordinate. (b): Paths of radiative heat exchange between a nonblack element and two black elements: (1) direct radiative heat exchange between black surfaces (black lines); (2) radiation emitted from $dA_i$ and reflected by nonblack element towards $dA_j$ (blue line); and (3) radiation emitted from nonblack element to $dA_j$ (green line). Radiation emitted from $dA_j$ to $dA$ is partially absorbed and reflected towards $dA_i$ (red line).

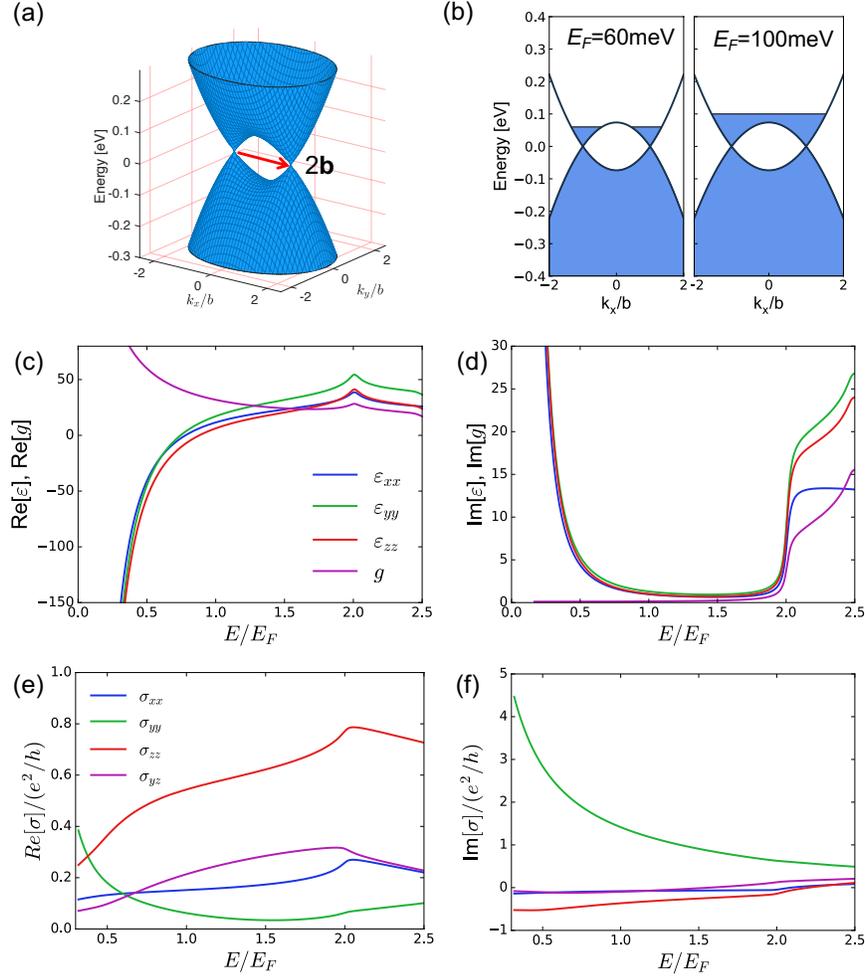

Figure 2: (a) Electronic band structure of the effective Hamiltonian on the surface of $k_z=0$, and (b) at the projection to the $k_y = k_z = 0$ plane. The occupied electron states are illustrated in the solid blue region. (c and e) Real and (d and f) imaginary parts of local dielectric tensor and surface optical conductivity of the magnetic Weyl semimetal modeled by the effective Hamiltonian for $E_F = 60$ meV. The Fermi velocity, the Weyl node separation, and the dissipative loss are $v_F = 1.0 \times 10^5$ m/s and $2b = 0.45$Å, and $\gamma = 1.5$meV, respectively.

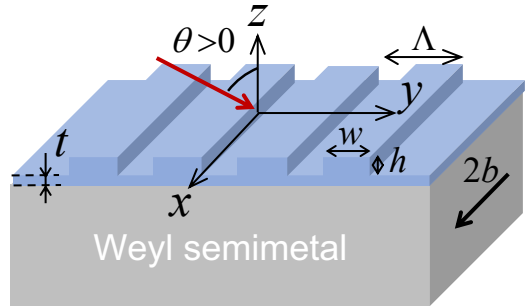

Figure 3: Optical grating structure made of a low-loss dielectric with $\varepsilon_d = 10 + i0.01$ on top of a semi-infinite magnetic Weyl semimetal with its Weyl node separation $2b$ in the $k_x$-direction. The period, width, height, and thickness of the gratings are $\Lambda$, $w$, $h$, and $t$, respectively. The light is incident at angle $\theta$ and $p$-polarized (the magnetic field is along the $x$-direction).

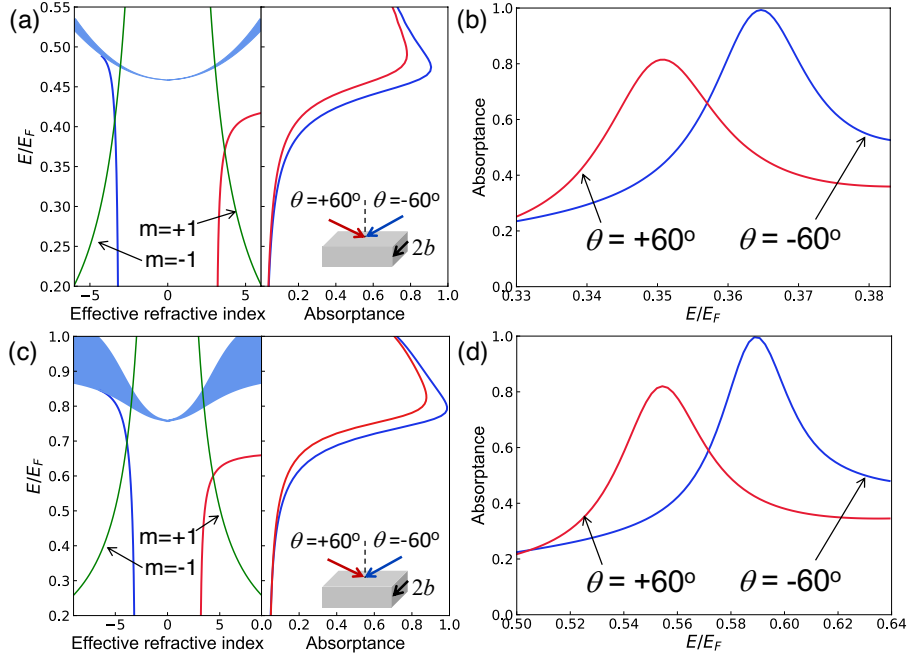

Figure 4: (a and c) Dispersion relation of SPPs for the angles of incidence $\theta = -60°$ (blue) and $\theta = +60°$ (red), respectively. The dispersion of the grating for the diffracting order $m = \pm 1$ (green) as well as that of bulk plasmons (blude region) are shown. With the shared $y$-axis, spectral absorptance of the semi-infinite Weyl semimetal for the incident angle $\theta = \pm 60°$ is shown. (b and d) Spectral directional absorptance of the grating structure for the incident angles of $\theta = \pm 60°$. All simulations do not include the Fermi-arc surface states. The upper (a and b) and lower (c and d) panels are for $E_F$= 60 meV and $E_F$=100 meV, respectively. The geometrical parameters of the grating structure are (b): $\Lambda = 20\mu m$, $w = \Lambda/2$, $h = 1.6\mu m$, $t = 3\mu m$, (d) $\Lambda = 5.9\ \mu m$, $w = \Lambda/2$, $h = 0.95\ \mu m$, $t = 1.2\ \mu m$.

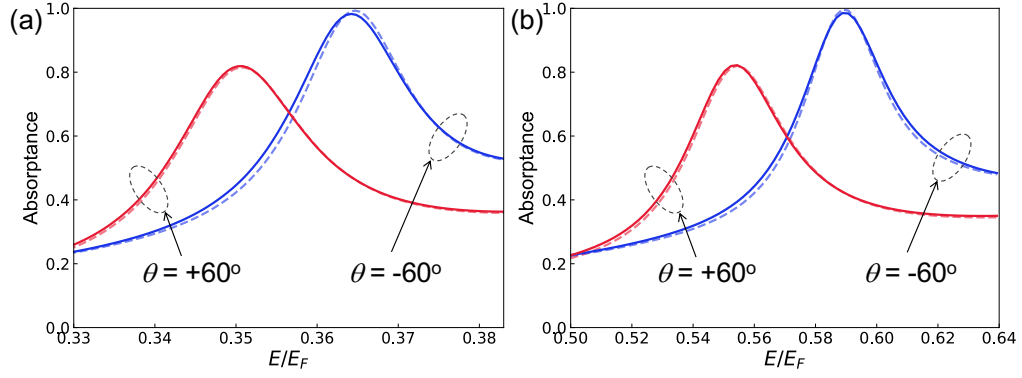

Figure 5: Spectral directional absorptance of the grating structure with (solid line) and without (dashed line) the presence of the Fermi-arc states for the incident angles of $\theta = \pm 60°$. The fermi energies are (a) $E_F$= 60 meV and (b) $E_F$=100 meV, respectively. The structure's geometrical parameters are the same as in Fig.4.

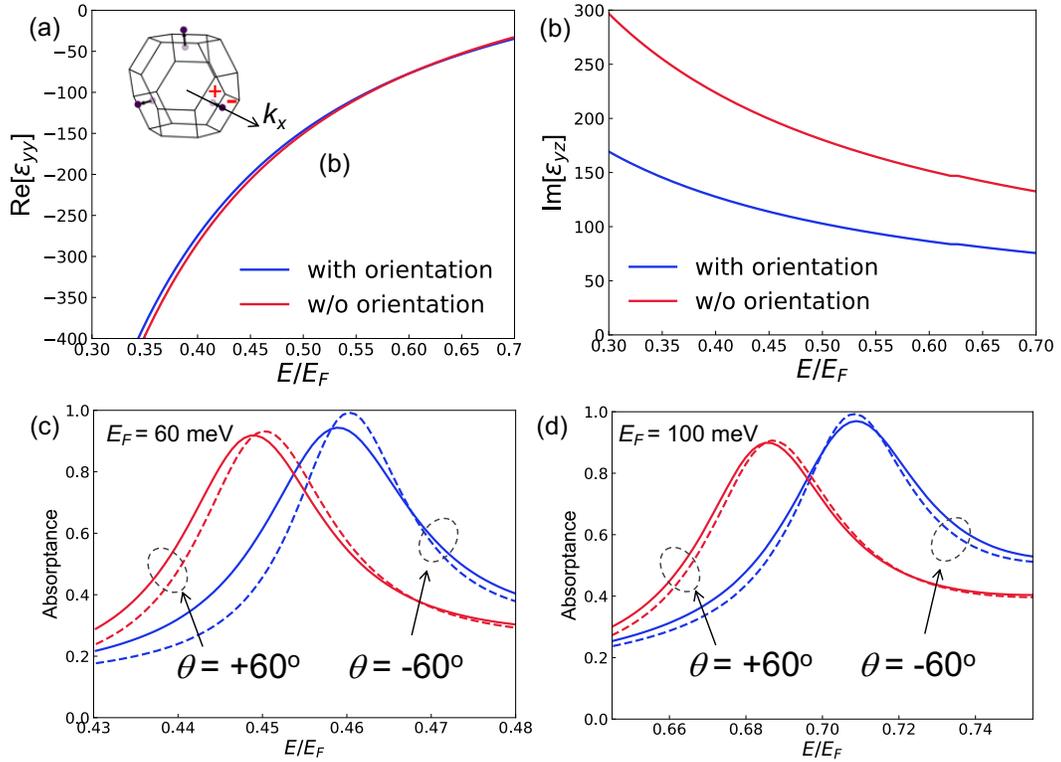

Figure 6: Comparison of (a) diagonal and (b) off-diagonal elements of dielectric tensor with and without the consideration of relative orientation of Weyl node pairs for $E_F$=60 meV. Inset of (a): Bulk Brillouin zone of rhombohedral cell and six Weyl nodes with positive and negative chirality. One of the pair of Weyl nodes is aligned along $k_x$-axis. (c and d) Spectral absorptance of the grating structure on the Weyl semimetal with three oriented Weyl node pairs with (solid line) and without (dashed line) the presence of the Fermi-arc states for the incident angles of $\theta = \pm 60°$. The Fermi energies are (c) 60 meV and (d) 100 meV, respectively. The geometrical parameters of the grating structure are (c): $\Lambda = 14.2 \mu m, w = \Lambda/1.5, h = 1.4 \mu m, t = 2.85 \mu m$, (d): $\Lambda = 5.3 \mu m, w = \Lambda/2, h = 0.7 \mu m, t = 1.0 \mu m$.

**Reference**

[1]  M. A. Green, Nano Lett. **12**, 5985 (2012).

[2]  L. Zhu and S. Fan, Phys. Rev. B **90**, 220301 (2014).

[3]  H. Ries, Appl. Phys. B Photophysics Laser Chem. **32**, 153 (1983).

[4]  Rayleigh, Nature **64**, 577 (1901).

[5]  W. C. Snyder, Z. Wan, and X. Li, Appl. Opt. **37**, 3464 (1998).

[6]  J.-J. Greffet and M. Nieto-Vesperinas, J. Opt. Soc. Am. A **15**, 2735 (1998).

[7]  J. J. Greffet, P. Bouchon, G. Brucoli, and F. Marquier, Phys. Rev. X **8**, 021008, (2018).

[8]  Y. Hadad, J. C. Soric, and A. Alu, Proc. Natl. Acad. Sci. U. S. A. **113**, 3471 (2016).

[9]  A. Zvezdin and V. A. Kotov, *Modern Magnetooptics and Magnetooptical Materials*, in (CRC Press, Boca Raton, FL, 1997).

[10] N. Nagaosa, J. Sinova, S. Onoda, A. H. MacDonald, and N. P. Ong, Rev. Mod. Phys. **82**, 1539 (2010).

[11] D. Xiao, M. C. Chang, and Q. Niu, Rev. Mod. Phys. **82**, 1959 (2010).

[12] V. I. Belotelov, I. A. Akimov, M. Pohl, V. A. Kotov, S. Kasture, A. S. Vengurlekar, A. V. Gopal, D. R. Yakovlev, A. K. Zvezdin, and M. Bayer, Nat. Nanotechnol. **6**, 370 (2011).

[13] G. Armelles, A. Cebollada, A. García-Martín, and M. U. González, Adv. Opt. Mater. **1**, 10

(2013).

[14]  Y. Shoji and T. Mizumoto, Sci. Technol. Adv. Mater. **15**, 014602 (2014).

[15]  B. J. H. Stadler and T. Mizumoto, IEEE Photonics J. **6**, 1 (2014).

[16]  L. Bi, J. Hu, P. Jiang, D. H. Kim, G. F. Dionne, L. C. Kimerling, and C. A. Ross, Nat. Photonics **5**, 758 (2011).

[17]  J. Y. Chin, T. Steinle, T. Wehlus, D. Dregely, T. Weiss, V. I. Belotelov, B. Stritzker, and H. Giessen, Nat. Commun. **4**, 1599 (2013).

[18]  B. Zhao, Y. Shi, J. Wang, Z. Zhao, N. Zhao, and S. Fan, Opt. Lett. **44**, 4203 (2019).

[19]  R. K. Hickernell and D. Sarid, Opt. Lett. **12**, 570 (1987).

[20]  N. P. Armitage, E. J. Mele, and A. Vishwanath, Rev. Mod. Phys. **90**, 015001 (2018).

[21]  X. Wan, A. M. Turner, A. Vishwanath, and S. Y. Savrasov, Phys. Rev. B **83**, 205101 (2011).

[22]  H. Weng, C. Fang, Z. Fang, B. Andrei Bernevig, and X. Dai, Phys. Rev. X **5**, 011029 (2015).

[23]  L. Wu, S. Patankar, T. Morimoto, N. L. Nair, E. Thewalt, A. Little, J. G. Analytis, J. E. Moore, and J. Orenstein, Nat. Phys. **13**, 350 (2017).

[24]  G. B. Osterhoudt, L. K. Diebel, M. J. Gray, X. Yang, J. Stanco, X. Huang, B. Shen, N. Ni, P. J. W. Moll, Y. Ran, and K. S. Burch, Nat. Mater. **18**, 471 (2019).

[25]  F. De Juan, A. G. Grushin, T. Morimoto, and J. E. Moore, Nat. Commun. **8**, 15995, (2017).

[26]  E. Liu, Y. Sun, N. Kumar, L. Muechler, A. Sun, L. Jiao, S. Y. Yang, D. Liu, A. Liang, Q. Xu, J. Kroder, V. Süß, H. Borrmann, C. Shekhar, Z. Wang, C. Xi, W. Wang, W. Schnelle, S. Wirth, Y. Chen, S. T. B. Goennenwein, and C. Felser, Nat. Phys. **14**, 1125 (2018).

[27]  Q. Wang, Y. Xu, R. Lou, Z. Liu, M. Li, Y. Huang, D. Shen, H. Weng, S. Wang, and H.

Lei, Nat. Commun. **9**, 3681 (2018).

[28] R. Singha, S. Roy, A. Pariari, B. Satpati, and P. Mandal, Phys. Rev. B **99**, 035110 (2019).

[29] N. J. Ghimire, A. S. Botana, J. S. Jiang, J. Zhang, Y. S. Chen, and J. F. Mitchell, Nat. Commun. **9**, 3280 (2018).

[30] A. K. Nayak, J. E. Fischer, Y. Sun, B. Yan, J. Karel, A. C. Komarek, C. Shekhar, N. Kumar, W. Schnelle, J. Kübler, C. Felser, and S. S. P. Parkin, Sci. Adv. **2**, 1501870 (2016).

[31] T. Miyasato, N. Abe, T. Fujii, A. Asamitsu, S. Onoda, Y. Onose, N. Nagaosa, and Y. Tokura, Phys. Rev. Lett. **99**, 086602, (2007).

[32] O. V. Kotov and Y. E. Lozovik, Phys. Rev. B **98**, 195446 (2018).

[33] R. Siegel and J. R. Howell, *Thermal Radiation Heat Transfer - Third Edition* (1992).

[34] M. F. Modest, *Radiative Heat Transfer* (Elsevier, Academic press, Amsterdam, 2013).

[35] F. J. J. Clarke and D. J. Parry, Light. Res. Technol. **17**, 1 (1985).

[36] J. Hofmann and S. Das Sarma, Phys. Rev. B **93**, 241402(R) (2016).

[37] F. M. D. Pellegrino, M. I. Katsnelson, and M. Polini, Phys. Rev. B **92**, 201407(R) (2015).

[38] A. A. Zyuzin and A. A. Burkov, Phys. Rev. B **86**, 115133 (2012).

[39] A. A. Burkov, Phys. Rev. Lett. **113**, 187202 (2014).

[40] C. A. C. Garcia, J. Coulter, and P. Narang, Phys. Rev. Research **2**, 013073, (2020).

[41] B. Xu, Y. M. Dai, L. X. Zhao, K. Wang, R. Yang, W. Zhang, J. Y. Liu, H. Xiao, G. F. Chen, A. J. Taylor, D. A. Yarotski, R. P. Prasankumar, and X. G. Qiu, Phys. Rev. B **93**, 121110(R) (2016).

[42] S. I. Kimura, H. Yokoyama, H. Watanabe, J. Sichelschmidt, V. Süß, M. Schmidt, and C. Felser, Phys. Rev. B **96**, 075119 (2017).

[43] Q. Chen, A. R. Kutayiah, I. Oladyshkin, M. Tokman, and A. Belyanin, Phys. Rev. B **99**,

075137 (2019).

[44] C. J. Tabert, J. P. Carbotte, and E. J. Nicol, Phys. Rev. B **93**, 085426 (2016).

[45] R. Okugawa and S. Murakami, Phys. Rev. B **89**, 235315 (2014).

[46] S. Murakami and S. I. Kuga, Phys. Rev. B **78**, 165313 (2008).

[47] See Supplemental Material for detailed information on modeling and calculation results of dielectric functions and derivation of electromagnetic interface conditions.

[48] I. Belopolski, K. Manna, D. S. Sanchez, G. Chang, B. Ernst, J. Yin, S. S. Zhang, T. Cochran, N. Shumiya, H. Zheng, B. Singh, G. Bian, D. Multer, M. Litskevich, X. Zhou, S. M. Huang, B. Wang, T. R. Chang, S. Y. Xu, A. Bansil, C. Felser, H. Lin, and M. Zahid Hasan, Science (80-. ). **365**, 1278 (2019).

[49] F. Tang, H. C. Po, A. Vishwanath, and X. Wan, Nature **566**, 486 (2019).

[50] M. G. Vergniory, L. Elcoro, C. Felser, N. Regnault, B. A. Bernevig, and Z. Wang, Nature **566**, 480 (2019).

[51] T. Zhang, Y. Jiang, Z. Song, H. Huang, Y. He, Z. Fang, H. Weng, and C. Fang, Nature **566**, 475 (2019).

[52] P. Hosur, S. A. Parameswaran, and A. Vishwanath, Phys. Rev. Lett. **108**, 046602 (2012).

[53] A. B. Sushkov, J. B. Hofmann, G. S. Jenkins, J. Ishikawa, S. Nakatsuji, S. Das Sarma, and H. D. Drew, RAPID Commun. Phys. Rev. B **92**, 241108 (2015).

[54] Ž. Bonačić Losic, J. Phys. Condens. Matter **30**, 285001, (2018).

[55] J. C. W. Song and M. S. Rudner, Phys. Rev. B **96**, 205443, (2017).

[56] G. M. Andolina, F. M. D. Pellegrino, F. H. L. Koppens, and M. Polini, Phys. Rev. B **97**, 125431, (2018).

[57] J. Zhou, H. R. Chang, and D. Xiao, Phys. Rev. B - Condens. Matter Mater. Phys. **91**,

035114, (2015).

[58] K. W. Chiu and J. J. Quinn, Nuovo Cim. B **10**, 1 (1972).

[59] B. Wunsch, T. Stauber, F. Sols, and F. Guinea, New J. Phys. **8**, (2006).

[60] X. T. Ji, H. Z. Lu, Z. G. Zhu, and G. Su, AIP Adv. **7**, 105003 (2017).

[61] T. Nguyen, F. Han, N. Andrejevic, R. Pablo-Pedro, A. Apte, Y. Tsurimaki, Z. Ding, K. Zhang, A. Alatas, E. E. Alp, S. Chi, J. Fernandez-Baca, M. Matsuda, D. A. Tennant, Y. Zhao, Z. Xu, J. W. Lynn, S. Huang, and M. Li, ArXiv:1906.00539 (2019).

[62] D. F. Liu, A. J. Liang, E. K. Liu, Q. N. Xu, Y. W. Li, C. Chen, D. Pei, W. J. Shi, S. K. Mo, P. Dudin, T. Kim, C. Cacho, G. Li, Y. Sun, L. X. Yang, Z. K. Liu, S. S. P. Parkin, C. Felser, and Y. L. Chen, Science (80-. ). **365**, 1282 (2019).

[63] W. Shi, L. Muechler, K. Manna, Y. Zhang, K. Koepernik, R. Car, J. Van Den Brink, C. Felser, and Y. Sun, Phys. Rev. B **97**, 060406(R) (2018).

[64] H. Ido and S. Yasuda, Le J. Phys. Colloq. **49**, C8 (1988).
30

# Supplementary Information on

# Large nonreciprocal absorption and emission of radiation

# in type-I Weyl semimetals with time reversal symmetry breaking

Yoichiro Tsurimaki[1,†], Xin Qian[1,†], Simo Pajovic[1], Fei Han[2], Mingda Li[2] and Gang Chen[1,*]

[1]Department of Mechanical Engineering, Massachusetts Institute of Technology, Cambridge, MA, 02139, USA

[2]Department of Nuclear Science and Engineering, Massachusetts Institute of Technology, Cambridge, MA, 02139, USA

## 1. Semi-analytical expressions of the bulk dielectric and surface conductivity tensors

In this work, the bulk dielectric tensor and the surface conductivity tensors are based on the work by Chen et al. [1]. In this section, we briefly review the semi-analytical expressions used in this work. The detailed derivation is given in [1]. The eigenstates of the effective Hamiltonian $|m\rangle$ can be analytically derived and the bulk optical conductivity is calculated by the Kubo formula:

$$\sigma_{\alpha\beta}(\omega) = \frac{gi\hbar}{V} \sum_{mn} \left(\frac{f_n - f_m}{E_m - E_n}\right) \frac{\langle n|\hat{j}_\alpha|m\rangle\langle m|\hat{j}_\beta|n\rangle}{\hbar\omega + E_n - E_m + \hbar\gamma}, \qquad (S1.1)$$

where $V$ is the volume of the material, $g$ is the spin degeneracy, $f$ is the Fermi-Dirac distribution, $E$ is the eigenenergy, $\gamma$ is the inverse of the mode lifetime, and $\hat{j}_\alpha$ is the current density operator in $\alpha$ direction ($\alpha = x, y, z$). Once the optical conductivity is known, the bulk dielectric tensor is calculated in Gauss unit by:

$$\varepsilon_{ij} = \varepsilon_b \delta_{ij} + \frac{4\pi}{\omega}\sigma_{ij}(\omega), \qquad (S1.2)$$

[*]Author to whom correspondence should be addressed. E-mail: gchen2@mit.edu, [†]These authors contributed equally.

At zero temperature, the bulk optical conductivities in Gauss unit in different directions have semi-analytical expressions as follows:

i) **Intra-band transition**

$$\sigma_{xx}^{intra}(\omega) = \frac{ige^2 v_F^2}{4\pi^3 b^2 k_F \hbar(\omega+i\gamma)} \int_{-\infty}^{\infty} dk_x \int_{-\infty}^{\infty} dk_y \frac{k_x^2 K_x^2 \Theta\left(k_F - \sqrt{K_x^2 + k_y^2}\right)}{\sqrt{k_F^2 - K_x^2 - k_y^2}}, \quad (S1.3)$$

$$\sigma_{yy}^{intra}(\omega) = \frac{ige^2 v_F^2}{4\pi^3 b^2 k_F \hbar(\omega+i\gamma)} \int_{-\infty}^{\infty} dk_x \int_{-\infty}^{\infty} dk_y \frac{k_y^2 (K_x + b)^2 \Theta\left(k_F - \sqrt{K_x^2 + k_y^2}\right)}{\sqrt{k_F^2 - K_x^2 - k_y^2}}, \quad (S1.4)$$

$$\sigma_{zz}^{intra}(\omega) = \frac{ige^2 v_F^2}{4\pi^3 b^2 k_F \hbar(\omega+i\gamma)} \int_{-\infty}^{\infty} dk_x \int_{-\infty}^{\infty} dk_y \, \Theta\left(k_F - \sqrt{K_x^2 + k_y^2}\right) \sqrt{k_F^2 - K_x^2 - k_y^2}, \quad (S1.5)$$

where $K_x = \frac{k_x^2 + k_y^2 - b^2}{2b}$ and $k_F = \frac{E_F}{\hbar v_F}$. $\Theta(x)$ is the Heaviside step function. The off-diagonal elements of the intraband transition are all zero:

$$\sigma_{xy}^{intra}(\omega) = \sigma_{yz}^{intra}(\omega) = \sigma_{zx}^{intra}(\omega) = 0. \quad (S1.6)$$

ii) **Inter-band transition**

$$\sigma_{xx}^{inter}(\omega) = \frac{ige^2(\omega+i\gamma)}{8\pi^3 b^2 \hbar v_F} \int_{-\infty}^{\infty} dk_x \int_{-\infty}^{\infty} dk_y \, 2k_x^2 [F_{xx}(k_F) - F_{xx}(K)], \quad (S1.7)$$

where the function $F_{xx}$ is

$$F_{xx}(x) = \Theta\left(x - \sqrt{K_x^2 + k_y^2}\right) \left( \frac{\frac{K_x^2 \sqrt{x^2 - K_x^2 - k_y^2}}{x \left(\frac{\omega+i\gamma}{v_F}\right)^2 (K_x^2 + k_y^2)}}{} + \frac{\left[\left(\frac{\omega+i\gamma}{v_F}\right)^2 - 4K_x^2\right] \tan^{-1}\left[\frac{\left(\frac{\omega+i\gamma}{v_F}\right)\sqrt{x^2 - K_x^2 - k_y^2}}{x\sqrt{4(K_x^2+k_y^2) - \left(\frac{\omega+i\gamma}{v_F}\right)^2}}\right]}{\left(\frac{\omega+i\gamma}{v_F}\right)^3 \sqrt{4(K_x^2+k_y^2) - \left(\frac{\omega+i\gamma}{v_F}\right)^2}} \right), \quad (S1.8)$$

and $K$ is a cutoff wavevector introduced to regularize the integral. In this work, we used the cutoff energy $E = \hbar v_F K = 2\text{eV}$. Other components are:

$$\sigma_{yy}^{inter}(\omega) = \frac{ige^2(\omega + i\gamma)}{4\pi^3 b^2 \hbar v_F} \int_{-\infty}^{\infty} dk_x \int_{-\infty}^{\infty} dk_y \left[F_{yy}(k_F) - F_{yy}(K)\right], \tag{S1.9}$$

where the function $F_{yy}$ is

$$F_{yy}(y) = \Theta\left(y - \sqrt{K_x^2 + k_y^2}\right) \times$$

$$\left( \frac{(K_x + b)^2 k_y^2 \sqrt{y^2 - K_x^2 - k_y^2}}{k_F \left(\frac{\omega + i\gamma}{v_F}\right)^2 (K_x^2 + k_y^2)} \right.$$

$$\left. + \frac{\left[\left(\frac{\omega + i\gamma}{v_F}\right)^2 (k_y^2 + b^2) - 4(K_x + b)^2 k_y^2\right] \tan^{-1}\left[\frac{\left(\frac{\omega + i\gamma}{v_F}\right)\sqrt{y^2 - K_x^2 - k_y^2}}{y\sqrt{4(K_x^2 + k_y^2) - \left(\frac{\omega + i\gamma}{v_F}\right)^2}}\right]}{\left(\frac{\omega + i\gamma}{v_F}\right)^3 \sqrt{4(K_x^2 + k_y^2) - \left(\frac{\omega + i\gamma}{v_F}\right)^2}} \right), \tag{S1.10}$$

$$\sigma_{zz}^{inter}(\omega) = \frac{ige^2(\omega + i\gamma)}{8\pi^3 \hbar v_F} \int_{-\infty}^{\infty} dk_x \int_{-\infty}^{\infty} dk_y \left[F_{zz}(K) - F_{zz}(k_F)\right], \tag{S1.11}$$

where the function $F_{zz}$ is

$$F_{zz}(z) = \Theta\left(z - \sqrt{K_x^2 + k_y^2}\right) \times$$

$$\left( \frac{2\sqrt{z^2 - K_x^2 - k_y^2}}{z\left(\frac{\omega + i\gamma}{v_F}\right)^2} \right.$$

$$\left. - \frac{8(K_x^2 + k_y^2) \tan^{-1}\left[\frac{\left(\frac{\omega + i\gamma}{v_F}\right)\sqrt{z^2 - K_x^2 - k_y^2}}{z\sqrt{4(K_x^2 + k_y^2) - \left(\frac{\omega + i\gamma}{v_F}\right)^2}}\right]}{\left(\frac{\omega + i\gamma}{v_F}\right)^3 \sqrt{4(K_x^2 + k_y^2) - \left(\frac{\omega + i\gamma}{v_F}\right)^2}} \right). \tag{S1.12}$$

The only non-zero off-diagonal element is $\sigma_{yz}^{inter}(\omega)$ and is given as:

$$\sigma_{yz}^{inter}(\omega) = -\frac{ge^2}{4\pi^3 b\hbar}\int_{-\infty}^{\infty}dk_x\int_{-\infty}^{\infty}dk_y\,(k_y^2 - bK_x) \times [F_{yz}(k_F) - F_{yz}(K)], \qquad (S1.13)$$

where the function $F_{yz}(y)$ is:

$$F_{yz}(y) = \Theta\left(y - \sqrt{K_x^2 + k_y^2}\right)\frac{2\tan^{-1}\left[\dfrac{\left(\dfrac{\omega+i\gamma}{v_F}\right)\sqrt{y^2 - K_x^2 - k_y^2}}{y\sqrt{4(K_x^2 + k_y^2) - \left(\dfrac{\omega+i\gamma}{v_F}\right)^2}}\right]}{\left(\dfrac{\omega+i\gamma}{v_F}\right)\sqrt{4(K_x^2 + k_y^2) - \left(\dfrac{\omega+i\gamma}{v_F}\right)^2}}. \qquad (S1.14)$$

The surface conductivity tensor is also calculated in a similar way and is composed of two transitions: surface-to-surface states transitions and surface-to-bulk states transitions. The expressions in Gauss unit are given below:

i) **Surface-to-surface states transitions**

$$\sigma_{yy}^{S,intra}(\omega) = \Theta(b - k_F)\frac{ige^2 v_F\sqrt{b^2 - k_F^2}}{2\pi^2 \hbar(\omega + i\gamma)}, \qquad (S1.15)$$

and all other components are zero.

ii) **Surface-to-bulk states transitions**

$$\sigma_{xx}^{S,inter}(\omega) = \frac{ige^2}{h}\int_0^\infty dk_z \int_{-\infty}^\infty dk_x \int_{-\infty}^\infty dk_y\, \Theta(b^2 - k_x^2 - k_y^2)\frac{k_z^2 k_x^2 K_x}{\pi^2(K_x^2 + k_z^2)^2 b^2} \times$$
$$\left[\frac{\Theta(k_F - \sqrt{K_x^2 + k_y^2 + k_z^2}) - \Theta(k_F + k_y)}{\sqrt{K_x^2 + k_y^2 + k_z^2}\left[\frac{\omega + i\gamma}{v_F} - k_y - \sqrt{K_x^2 + k_y^2 + k_z^2}\right]} - \frac{-\Theta(-k_F - k_y)}{\sqrt{K_x^2 + k_y^2 + k_z^2}\left[\frac{\omega + i\gamma}{v_F} - k_y + \sqrt{K_x^2 + k_y^2 + k_z^2}\right]}\right], \quad (S1.16)$$

$$\sigma_{yy}^{S,inter}(\omega) = \frac{ige^2}{h}\int_0^\infty dk_z \int_{-\infty}^\infty dk_x \int_{-\infty}^\infty dk_y\, \Theta(b^2 - k_x^2 - k_y^2)\frac{k_z^2 k_y^2 K_x}{\pi^2(K_x^2 + k_z^2)^2 b^2} \times$$
$$\left[\frac{\Theta(k_F - \sqrt{K_x^2 + k_y^2 + k_z^2}) - \Theta(k_F + k_y)}{\sqrt{K_x^2 + k_y^2 + k_z^2}\left[\frac{\omega + i\gamma}{v_F} - k_y - \sqrt{K_x^2 + k_y^2 + k_z^2}\right]} - \frac{-\Theta(-k_F - k_y)}{\sqrt{K_x^2 + k_y^2 + k_z^2}\left[\frac{\omega + i\gamma}{v_F} - k_y + \sqrt{K_x^2 + k_y^2 + k_z^2}\right]}\right], \quad (S1.17)$$

$$\sigma_{zz}^{S,inter}(\omega) = \frac{ige^2}{h}\int_0^\infty dk_z \int_{-\infty}^\infty dk_x \int_{-\infty}^\infty dk_y\, \Theta(b^2 - k_x^2 - k_y^2)\frac{k_z^2 K_x}{\pi^2(K_x^2 + k_z^2)^2} \times$$
$$\left[\frac{\Theta(k_F - \sqrt{K_x^2 + k_y^2 + k_z^2}) - \Theta(k_F + k_y)}{\sqrt{K_x^2 + k_y^2 + k_z^2}\left[\frac{\omega + i\gamma}{v_F} - k_y - \sqrt{K_x^2 + k_y^2 + k_z^2}\right]} - \frac{-\Theta(-k_F - k_y)}{\sqrt{K_x^2 + k_y^2 + k_z^2}\left[\frac{\omega + i\gamma}{v_F} - k_y + \sqrt{K_x^2 + k_y^2 + k_z^2}\right]}\right], \quad (S1.18)$$

and the only non-zero off-diagonal element is

$$\sigma_{yz}^{S,inter}(\omega) = -\frac{ge^2}{h}\int_0^\infty dk_z \int_{-\infty}^\infty dk_x \int_{-\infty}^\infty dk_y\, \Theta(b^2 - k_x^2 - k_y^2)\frac{k_z^2 k_y K_x}{\pi^2(K_x^2 + k_z^2)^2 b} \times$$
$$\left[\frac{\Theta(k_F - \sqrt{K_x^2 + k_y^2 + k_z^2}) - \Theta(k_F + k_y)}{\sqrt{K_x^2 + k_y^2 + k_z^2}\left[\frac{\omega + i\gamma}{v_F} - k_y - \sqrt{K_x^2 + k_y^2 + k_z^2}\right]} - \frac{-\Theta(-k_F - k_y)}{\sqrt{K_x^2 + k_y^2 + k_z^2}\left[\frac{\omega + i\gamma}{v_F} - k_y + \sqrt{K_x^2 + k_y^2 + k_z^2}\right]}\right]. \quad (S1.19)$$

## 2. Frequency dependence of anomalous optical conductivity

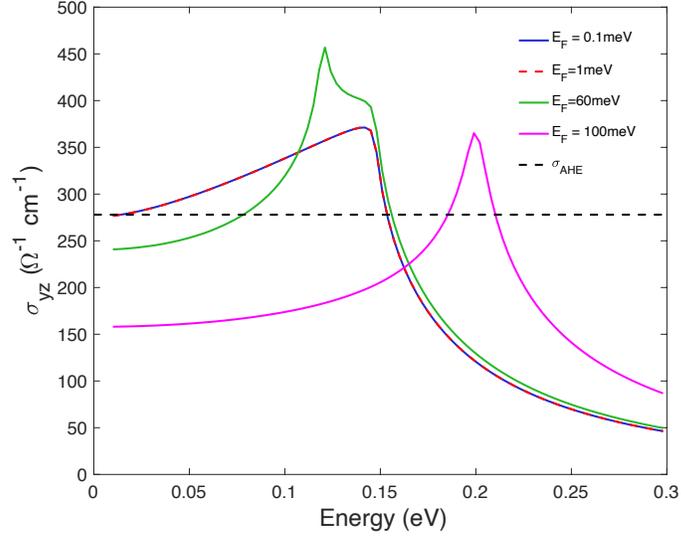

Figure S2.1: Frequency dependence of off-diagonal optical conductivity $\sigma_{yz}$ at different Fermi energies.

At low Fermi energies, the anomalous Hall conductivity from our model matches with the expression of the anomalous Hall conductivity of the Weyl semimetal with its Fermi energy at the Weyl node $\sigma_{AHE} = 2e^2 \frac{|\mathbf{b}|}{\pi\hbar} = 277 \, \Omega^{-1}\text{cm}^{-1}$ as shown in dashed black line. As long as the Fermi energy is below $E_F \lesssim 60\text{meV}$ so the two Fermi surfaces of the two Weyl nodes are separate, the anomalous Hall conductivity at low energy is close to $\sigma_{AHE}$. However, at Fermi energy 100 meV, the anomalous conductivity drastically decreases to $\sigma_{AHE} \sim 155 \, \Omega^{-1}\text{cm}^{-1}$. At finite frequency, the optical conductivity (ac conductivity) slightly increases above the dc conductivity.

## 3. Bulk dielectric tensor and surface conductivity tensor at $E_F = 100$ meV

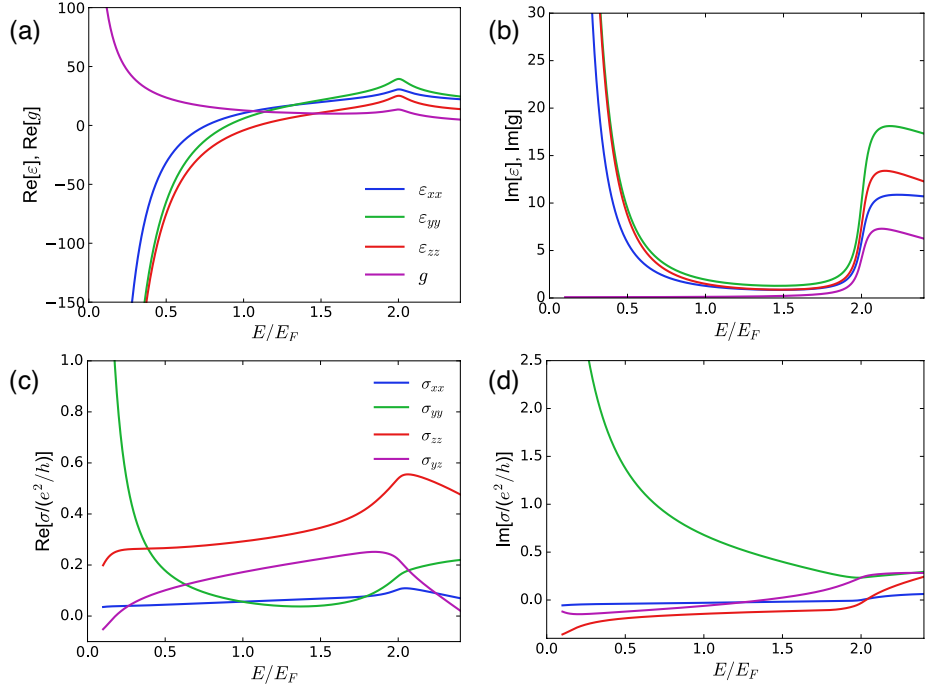

Figure S3.1: Real (a and c) and imaginary (b and d) parts of local bulk dielectric tensor and surface optical conductivity of a magnetic Weyl semimetal that possesses a single pair of Weyl nodes for $E_F$=100 meV. The Fermi velocity, the Weyl node separation, and the dissipative loss are assumed $v_F = 1.0 \times 10^5$ m/s and $2b = 0.45$Å, and $\gamma = 4.2$meV, respectively.

## 4. Electromagnetic interface conditions for Weyl semimetals

The derivation of the electromagnetic interface conditions follows the discussion in [1]. We consider an isotropic, non-dispersive dielectric material with a dielectric constant $\varepsilon_d$ at $z > 0$ and a magnetic Weyl semimetal at $z < 0$, forming an interface at $z = 0$. We derive the interface conditions of the electromagnetic field in the presence of the Fermi arc surface states.

Maxwell's equations in the frequency domain when we use the time-harmonic field $e^{-i\omega t}$ are:

$$\nabla \times \boldsymbol{E}(\boldsymbol{x}, \omega) - i\omega \boldsymbol{B}(\boldsymbol{x}, \omega) = 0, \tag{S4.1}$$

$$\nabla \times \boldsymbol{H}(\boldsymbol{x}, \omega) + i\omega \boldsymbol{D}(\boldsymbol{x}, \omega) = \boldsymbol{j}_e^S(\boldsymbol{x}, \omega), \tag{S4.2}$$

$$\nabla \cdot \boldsymbol{D}(\boldsymbol{x},\omega) = \rho^S(\boldsymbol{R},\omega)\delta(z-0^-), \tag{S4.3}$$

$$\nabla \cdot \boldsymbol{B}(\boldsymbol{x},\omega) = 0. \tag{S4.4}$$

Due to the presence of the Fermi arc surface states, we assume the surface charge density $\rho^S(\boldsymbol{R},\omega)\delta(z-0^-)$ just beneath the surface of the Weyl semimetal. As a result, there exists a current density $\boldsymbol{j}_e^S(\boldsymbol{x},\omega)$. $\boldsymbol{R}=(x,y)$ is the coordinate in the *x-y* plane and $\boldsymbol{x}=(\boldsymbol{R},z)$. Note that the delta function has units of inverse length, with which the units of the surface charge density become Coulomb/m³. The unit of the current density is Ampere/m². The electric induced displacement fields are $\boldsymbol{D}(\boldsymbol{x},\omega) = \varepsilon_0\varepsilon_d \boldsymbol{E}(\boldsymbol{x},\omega)$ in the dielectric and $\boldsymbol{D}(\boldsymbol{x},\omega) = \varepsilon_0(\varepsilon_b + i\hat{\sigma}^B(\omega)/\varepsilon_0\omega)\boldsymbol{E}(\boldsymbol{x},\omega)$ in the bulk magnetic Weyl semimetal, where $\varepsilon_0$ is the vacuum permittivity and $\varepsilon_b$ accounts for the free-carrier response from other bands besides the Weyl cones as well as the dielectric response. $\hat{\sigma}^B(\omega)$ is the bulk electrical conductivity tensor due to the bulk Weyl fermions. By definition, the relation between the parallel components of the surface current and the current density is:

$$\boldsymbol{i}_e^S(\boldsymbol{R},\omega) = \int_{-L}^{0} \boldsymbol{j}_e^S(\boldsymbol{x},\omega)dz, \tag{S4.5}$$

where $L$ is a length over which the charge flow exists near the surface.

Like the bulk conductivity, the surface current also follows the form of Ohm's law:

$$\boldsymbol{i}_e^S(\boldsymbol{R},\omega) = \hat{\sigma}^S(\omega)\boldsymbol{E}(\boldsymbol{x},z=0^-,\omega), \tag{S4.6}$$

where $\hat{\sigma}^S(\omega)$ is the surface optical conductivity tensor. The electric field is evaluated just beneath the surface of the Weyl semimetal at $z = 0^-$. The surface conductivity tensor due to the Fermi arc states has the same matrix form as the bulk conductivity:

$$\hat{\sigma}^S(\omega) = \begin{bmatrix} \sigma_{xx}^S(\omega) & 0 & 0 \\ 0 & \sigma_{yy}^S(\omega) & \sigma_{yz}^S(\omega) \\ 0 & -\sigma_{yz}^S(\omega) & \sigma_{zz}^S(\omega) \end{bmatrix}. \tag{S4.7}$$

The non-zero elements $\sigma_{zz}^S(\omega)$ and $\sigma_{yz}^S(\omega)$ due to the transition between the surface and bulk states create the non-zero surface current normal to the surface $i_{e,z}^S(\boldsymbol{R},\omega)$. In the interface conditions, we must reflect all the three components of the surface current.

From Eqs. (S4.3) and (S4.4), we will obtain the interface conditions about the normal components of the electric displacement field and the magnetic induction field while Eqs. (S4.1) and (S4.2) will give the interface conditions about their tangential components. Two of the four interface conditions are independent. Here we use the interface conditions about the tangential components. By integrating Eq. (S4.2) over a thin rectangular with height $2L$ across the interface and width $w$ on the $x$-$z$ plane so that the center of the rectangular is on $z = 0$, we have:

$$\int d\boldsymbol{A} \cdot [\nabla \times \boldsymbol{H}(\boldsymbol{x},\omega) + i\omega \boldsymbol{D}(\boldsymbol{x},\omega)] = \underbrace{\int d\boldsymbol{A} \cdot \boldsymbol{j}_e^S(\boldsymbol{x},\omega)}_{-i_{e,y}^S(\boldsymbol{R},\omega)w}. \quad (S4.8)$$

Noting that $d\boldsymbol{A} = -\boldsymbol{e}_y dA$ where $\boldsymbol{e}_y$ is the unit vector in $y$-direction, this equation becomes

$$-(H_x(\boldsymbol{R}, z = L, \omega) - H_x(\boldsymbol{R}, z = -L, \omega))w + \int d\boldsymbol{A} \cdot i\omega \boldsymbol{D}(\boldsymbol{x},\omega) = -i_{e,y}^S(\boldsymbol{R},\omega)w, \quad (S4.9)$$

where we used Eq.(S4.5) for the right hand side. By taking $L$ to be small, we obtain the interface condition for $H_x$:

$$H_x(\boldsymbol{R}, z = 0^+, \omega) - H_x(\boldsymbol{R}, z = 0^-, \omega) = i_{e,y}^S(\boldsymbol{R},\omega) \quad (S4.10)$$

Similarly, we obtain the interface condition for $H_y$ by integrating over the $y$-$z$ plane:

$$H_y(\boldsymbol{R}, z = 0^+, \omega) - H_y(\boldsymbol{R}, z = 0^-, \omega) = -i_{e,x}^S(\boldsymbol{R},\omega). \quad (S4.11)$$

Eqs. (S4.10) and (S4.11) are the interface condition about the tangential components of magnetic fields.

For the interface conditions of the tangential components of the electric field, we must reflect the fact that the surface current has a non-zero normal component $i_{e,z}^S(\boldsymbol{R},\omega)$. This fact, as we will see below, creates the dipole layer at the boundary of the media. The interface conditions about the tangential components of electric fields can be derived from Eq. (S4.1). Since we consider a semi-infinite and translationally invariant surface along $x$ and $y$ directions, we first Fourier transform Eq. (S4.1) about $\mathbf{R}$ by plane waves by using $\partial/\partial x \to ik_x$ and $\partial/\partial y \to ik_y$:

$$ik_y E_z - \frac{\partial}{\partial z}E_y = i\omega B_x, \quad \frac{\partial}{\partial z}E_x - ik_x E_z = i\omega B_y, \tag{S4.13}$$

The relation of the tangential components of the electric field across the interface is given by the spatial derivative term $\frac{\partial}{\partial z}E$. By integrating both sides in Eq. (S4.13) over a length $2L$ across the interface, $\int_{-L}^{L} dz$, we have for the first equation in Eq. (S4.13):

$$ik_y \int_{-L}^{L} E_z(\mathbf{K},z,\omega)dz - E_y(\mathbf{K},z=L,\omega) + E_y(\mathbf{K},z=-L,\omega) = i\omega \int_{-L}^{L} B_x(\mathbf{K},z,\omega)dz \tag{S4.14}$$

where $\mathbf{K} = (k_x, k_y)$, and we used the translational invariance in order to take $k_y$ out of the integral. We assume $\omega L/c \ll 1$ where $c$ is the speed of light since we can always choose a sufficiently small $L$ for a given wavevector in vacuum $\omega/c$. Then we can neglect the magnetic induction field term and obtain:

$$E_y(\mathbf{K},z=L,\omega) - E_y(\mathbf{K},z=-L,\omega) = ik_y \int_{-L}^{L} E_z(\mathbf{K},z,\omega)dz. \tag{S4.15}$$

Before evaluating Eq. (S4.15), we consider the Gauss' law Eq. (S4.3) in the Fourier domain:

$$ik_x D_x(\mathbf{K},z,\omega) + ik_y D_y(\mathbf{K},z,\omega) + \frac{\partial}{\partial z}D_z(\mathbf{K},z,\omega) = \rho^S(\mathbf{K},\omega)\delta(z-0^-). \tag{S4.16}$$

By assuming $L$ is also sufficiently small so that the field variation in the $x$- and $y$-directions can be negligible compared to the variation in the $z$-direction, $|k_x|L \ll 1$ and $|k_y|L \ll 1$, we have:

$$\frac{\partial}{\partial z}D_z(\mathbf{K},z,\omega) = \rho^S(\mathbf{K},\omega)\delta(z-0^-). \tag{S4.17}$$

We aim to express the interface conditions in terms of the surface current, thus we use the charge density conservation law to write the charge density $\rho^S(\mathbf{K},\omega)\delta(z-0^-)$ in Eq. (S4.17) in terms of the surface current. The charge density conservation law derived from Eqs. (S4.2) and (S4.3) in the time domain is:

$$\frac{\partial}{\partial t}\left(\rho^S(\mathbf{R},t)\delta(z-0^-)\right) + \nabla \cdot \mathbf{j}_e^S(\mathbf{R},t) = 0. \tag{S4.18}$$

In the Fourier transformed domain:

$$\begin{aligned}&-i\omega\rho^S(\mathbf{K},\omega)\delta(z-0^-)\\&+ik_x j_{e,x}^S(\mathbf{K},z,\omega) + ik_y j_{e,y}^S(\mathbf{K},z,\omega) + \frac{\partial}{\partial z}\left(j_{e,z}^S(\mathbf{K},z,\omega)\right) = 0\end{aligned} \tag{S4.19}$$

Under $|k_x|L \ll 1$ and $|k_y|L \ll 1$,

$$-i\omega\rho^S(\mathbf{K},z,\omega)\delta(z-0^-) + \frac{\partial}{\partial z}\left(j_{e,z}^S(\mathbf{K},z,\omega)\right) = 0. \tag{S4.20}$$

Inserting Eq. (S.20) into Eq. (S.17), we obtain:

$$\frac{\partial}{\partial z}D_z(\mathbf{K},z,\omega) = -\frac{i}{\omega}\frac{\partial}{\partial z}\left(j_{e,z}^S(\mathbf{K},z,\omega)\right). \tag{S4.21}$$

thereby we obtain $D_z(\mathbf{K},z,\omega) = -i/\omega j_{e,z}^S(\mathbf{K},z,\omega)$. Since the electric field in Eq. (S4.15) originates from the Faraday's law Eq. (S4.1) and it is about the electric field in vacuum, we separate the electric displacement field in Eq. (S4.21) into the vacuum field and the induced bulk material field $P_z$:

$$\varepsilon_0 E_z(\mathbf{K},z,\omega) = -P_z(\mathbf{K},z,\omega) - i/\omega\, j_{e,z}^S(\mathbf{K},z,\omega). \tag{S4.22}$$

Using this equation into Eq. (S4.15), we obtain:

$$E_y(\mathbf{K},z=L,\omega) - E_y(\mathbf{K},z=-L,\omega) = -\frac{ik_y}{\varepsilon_0}\int_{-L}^{L}(P_z + \frac{i}{\omega}j_{e,z}^S(K,z,\omega))dz. \tag{S4.23}$$

In small $L$ limit, the bulk polarization term vanishes, we obtain:

$$E_y(\boldsymbol{K}, z = L, \omega) - E_y(\boldsymbol{K}, z = -L, \omega) = \frac{k_y}{\varepsilon_0 \omega} \int_{-L}^{L} j_{e,z}^S(K, z, \omega)) dz$$

$$= \frac{k_y}{\varepsilon_0 \omega} \int_{-L}^{0} j_{e,z}^S(K, z, \omega)) dz, \quad (S4.24)$$

where in the last equality, we used the fact that free surface current does not exist inside the dielectric where $z > 0$. By definition, the current density flowing through the control volume with thickness $L$ is the dipole generation rate [2], thereby we have:

$$-i\omega \int_{-L}^{L} p_z dz = \int_{-L}^{L} j_{e,z}^S(\boldsymbol{x}, \omega) dz = -i\omega d_z, \quad (S4.25)$$

where $p_z$ is the polarization near the surface ($-L < z < 0$) and $d_z$ is the dipole moment. Therefore, we obtain the interfacial condition about the tangential components of the electric field as:

$$E_y(\boldsymbol{K}, z = 0^+, \omega) - E_y(\boldsymbol{K}, z = 0^-, \omega) = -i\frac{k_y}{\varepsilon_0} d_z(\boldsymbol{K}, z = 0^-, \omega). \quad (S4.26)$$

Similarly, for the *x*-component:

$$E_x(\boldsymbol{K}, z = 0^+, \omega) - E_x(\boldsymbol{K}, z = 0^-, \omega) = -i\frac{k_x}{\varepsilon_0} d_z(\boldsymbol{K}, z = 0^-, \omega). \quad (S4.27)$$

The dipole moment is related to the surface current through:

$$d_z = \frac{i}{\omega} \int_{-L}^{L} j_{e,z}^S(\boldsymbol{x}, \omega) dz = \frac{i}{\omega} i_{e,z}^S(\boldsymbol{K}, \omega)$$

$$= \frac{i}{\omega} (\sigma_{zy}^S(\omega) E_y(\boldsymbol{K}, z = 0^-, \omega) + \sigma_{zz}^S(\omega) E_z(\boldsymbol{K}, z = 0^-, \omega)). \quad (S4.28)$$

Eqs. (S4.26) and (S4.27) are the interface conditions for the tangential components of the electric field.

We solve the Maxwell's equations with the finite-element method in the frequency domain, thus we express the interface conditions Eqs. (S4.26), (S4.27) in the real space:

$$E_y(\mathbf{R}, z=0^+, \omega) - E_y(\mathbf{R}, z=0^-, \omega)$$
$$= \frac{-i}{\varepsilon_0 \omega}\left(\sigma^S_{zy}\frac{\partial}{\partial y}E_y(\mathbf{R}, z=0^-, \omega) + \sigma^S_{zz}\frac{\partial}{\partial y}E_z(\mathbf{R}, z=0^-, \omega)\right), \quad \text{(S4.29)}$$

$$E_x(\mathbf{R}, z=0^+, \omega) - E_x(\mathbf{R}, z=0^-, \omega)$$
$$= \frac{-i}{\varepsilon_0 \omega}\left(\sigma^S_{zy}\frac{\partial}{\partial x}E_y(\mathbf{R}, z=0^-, \omega) + \sigma^S_{zz}\frac{\partial}{\partial x}E_z(\mathbf{R}, z=0^-, \omega)\right). \quad \text{(S4.30)}$$

Note that Eqs. (S4.29) and (S4.30) are the particular results when we assume the form of the surface conductivity tensor in Eq. (S4.7). When all elements of the surface optical conductivity tensor are non-zero, we must consider additional terms in Eqs. (S4.29) and (S4.30) which are straightforward to obtain.

## 5. Effects of grating parameters on spectral directional absorptance

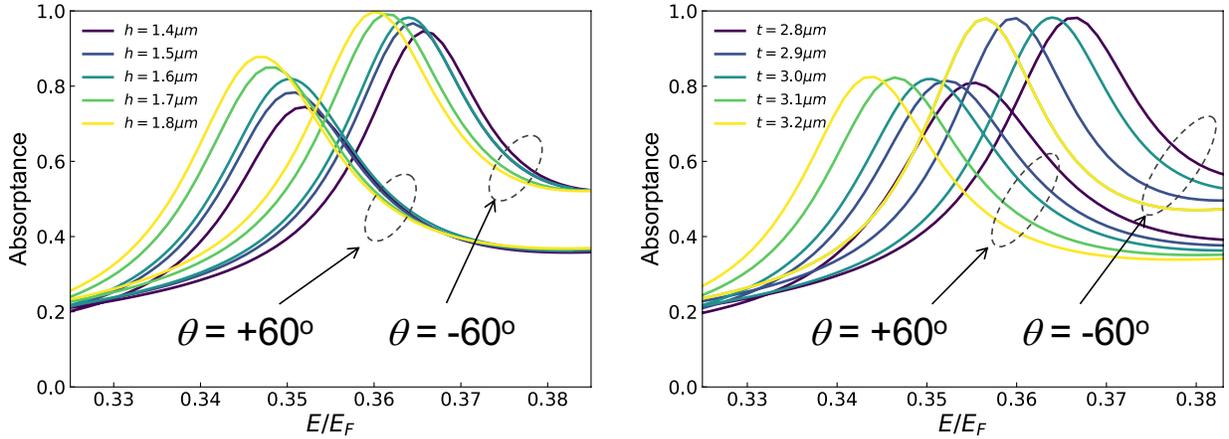

Figure S5.1: Spectral directional absorptance of the grating structure for the incident angles of $\theta = \pm 60°$. All simulations include the Fermi-arc surface states. The left panel (a) varies the grating parameter $h$ while the other grating parameters are fixed to $\Lambda = 20\mu m$, $w = \Lambda/2$, $t = 3.0\ \mu m$. The right panel (b) varies the grating parameter $t$ while fixing the other grating parameters to $\Lambda = 20\mu m$, $w = \Lambda/2$, $h = 1.6\mu m$. The Fermi energy is $E_F = 60$ meV.

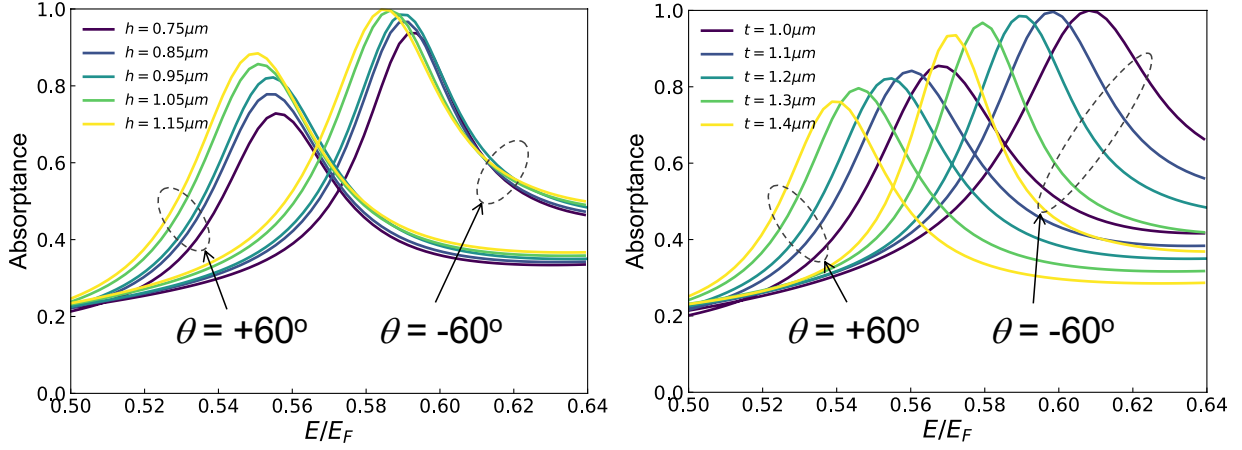

Figure S5.2: Spectral directional absorptance of the grating structure for the incident angles of $\theta=\pm60°$. All simulations include the Fermi-arc surface states. The left panel (a) varies the grating parameter $h$ while the other grating parameters are fixed to $\Lambda = 5.9\mu m, w = \Lambda/2, t = 1.2\ \mu m$. The right panel (b) varies the grating parameter $t$ while fixing the other grating parameters to $\Lambda = 5.9\mu m, w = \Lambda/2, h = 0.95\mu m$. The Fermi energy is $E_F = 100$ meV.

## 6. Effects of the orientation of Weyl node separation on optical responses

In realistic Weyl semimetals, multiple Weyl node pairs are not oriented in the same direction. To investigate the effects of the orientation on the dielectric and surface conductivity tensors and the spectral absorptance, we consider the relative orientation of the Weyl nodes found in $Co_3Sn_2S_2$ as a prototypical Weyl semimetal. There are a total six Weyl nodes and they are numerically calculated [3]. We model that the three pairs of Weyl node contribute independently to the overall dielectric and surface conductivity tensor and do not consider the inter-Weyl-node-pair interaction. In our model, the absolute locations of the Weyl nodes do not matter but only their relative orientations are reflected in the model.

The location of a Weyl point with positive chirality was found in the fraction coordinate of rhombohedral primitive cell as $(k_a, k_b, k_c) = (0.360922, -0.059795, -0.059809)$ [3]. The locations of the other five Weyl points can be determined by $C_3$ rotational symmetry and inversion symmetry of the rhombohedral cell. Let $\boldsymbol{b}_1$ be the separation vector of one of the Weyl pair at $(0.360922, -0.059795, -0.059809)$ and $(-0.360922 + 1, 0.059795, 0.059809)$ oriented from the positive to negative chirality. We transform to the orthonormal coordinate in which the $k_x$ axis is aligned with $\boldsymbol{b}_1$ via the linear transformation as:

$$\begin{bmatrix} k_x \\ k_y \\ k_z \end{bmatrix} = \begin{bmatrix} b & b\cos\beta & b\cos\beta \\ 0 & b\sin\beta & \dfrac{b\cos\beta\,(1 - \cos\beta)}{\sin\beta} \\ 0 & 0 & \dfrac{b(1 - 3\cos^2\beta + 2\cos^3\beta)^{0.5}}{\sin\beta} \end{bmatrix} \begin{bmatrix} k_a \\ k_b \\ k_c \end{bmatrix}, \quad (S5.1)$$

where $b = 1.433\,\text{Å}^{-1}$ and $\beta = 109.5°$ are the norm and separation angle of the reciprocal vectors in the fractional coordinate. In these orthonormal coordinates, the Weyl node pair separation $\boldsymbol{b}_1$ is along the $k_x$-axis. By the same transformation to the orthonormal coordinate, the dielectric function of the other two Weyl node pairs that are oriented in $\boldsymbol{b}_2$ and $\boldsymbol{b}_3$ directions can be written in the orthonormal coordinate in which $\boldsymbol{b}_1$ is aligned to the $k_x$-axis via the application of rotation matrix $R_2$ and $R_3$. The total dielectric tensor is written as:

$$\hat{\varepsilon}(\omega) = \varepsilon_b + \hat{\varepsilon}_{Weyl}(\omega) + R_2\hat{\varepsilon}_{Weyl}(\omega)(R_2)^T + R_3\hat{\varepsilon}_{Weyl}(\omega)(R_3)^T, \quad (S5.2)$$

where $\hat{\varepsilon}_{Weyl}(\omega)$ is the dielectric tensor of a single Weyl pair excluding the background contribution. All the elements of the total dielectric tensor are non-zero. The total surface conductivity tensor is also determined in the same manner:

$$\hat{\sigma}^S(\omega) = \hat{\sigma}^S{}_{Weyl}(\omega) + R_2\hat{\sigma}^S{}_{Weyl}(R_2)^T + R_3\hat{\sigma}^S{}_{Weyl}(R_3)^T. \quad (S5.3)$$

Figure S6.1-S6.4 show the real and imaginary parts of the elements of the bulk dielectric tensor calculated from Eqs. (S5.2) and (S5.3) for $E_F = 60$ meV. The dielectric tensor as well as the surface conductivity tensor can also be calculated for $E_F = 100$ meV in a similar manner.

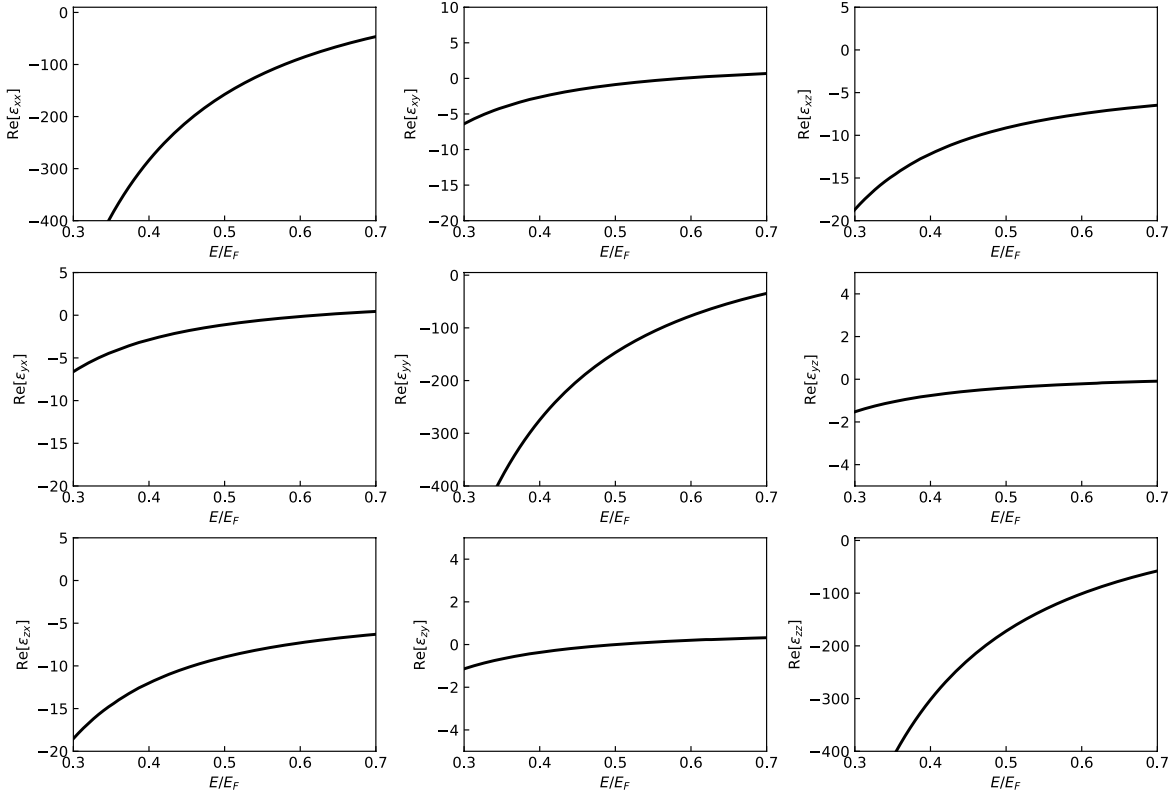

Figure S6.1: Real parts of the elements of the bulk dielectric tensor of a magnetic Weyl semimetal that possesses three oriented Weyl node pairs at $E_F = 60$ meV.

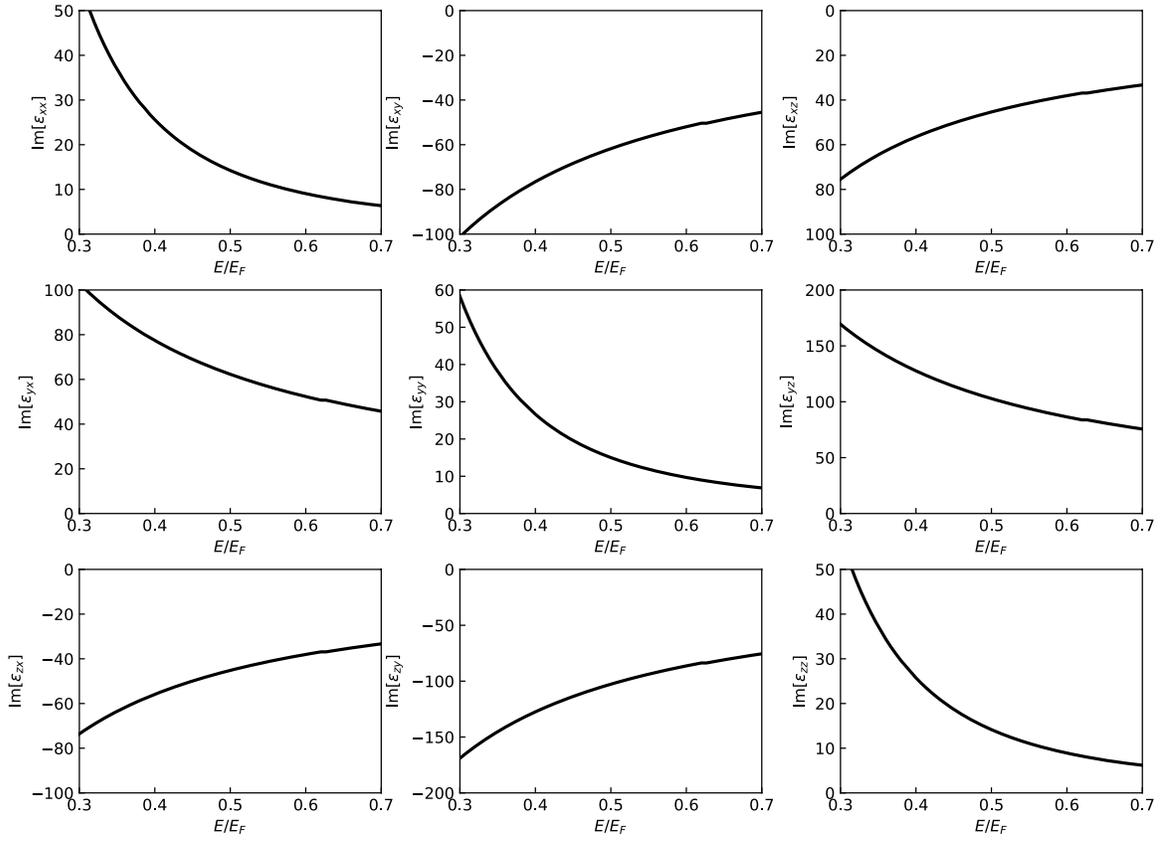

Figure S6.2: Imaginary parts of the elements of the bulk dielectric tensor of a magnetic Weyl semimetal that possesses three oriented Weyl node pairs at $E_F = 60$ meV.

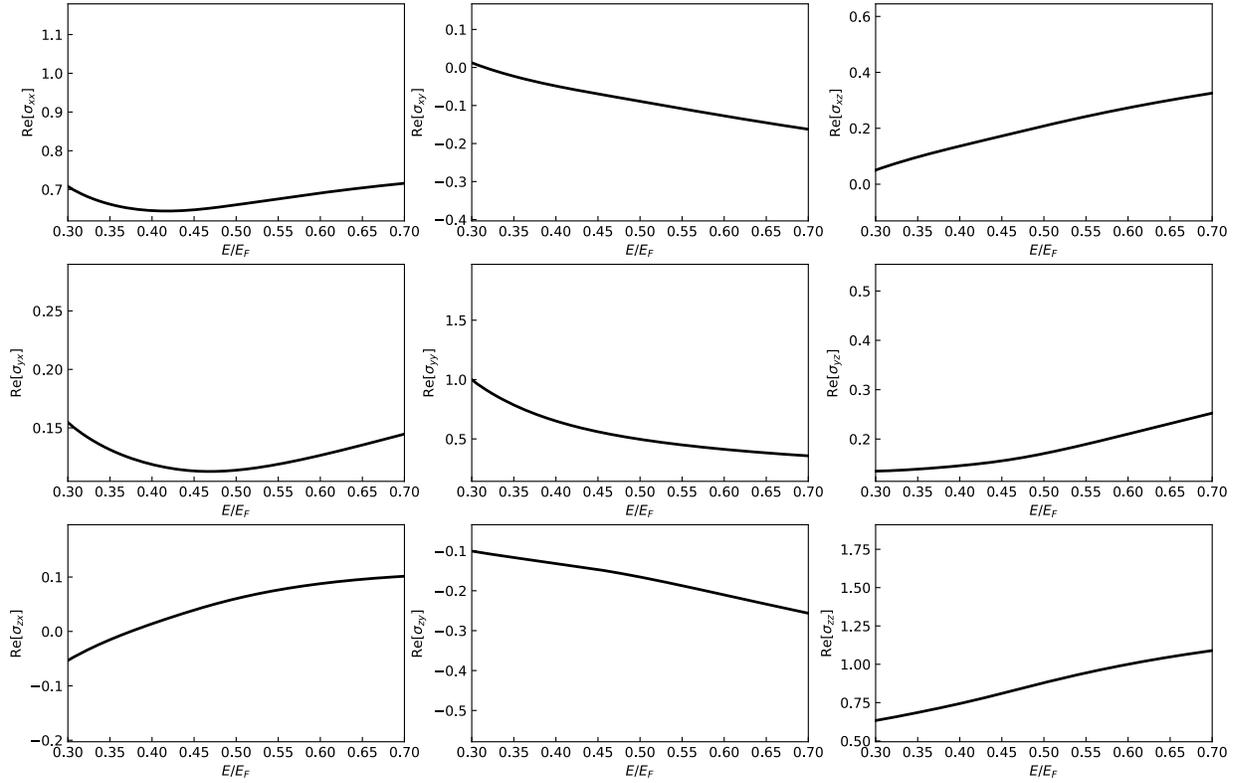

Figure S6.3: Real parts of the elements of the surface conductivity tensor of a magnetic Weyl semimetal that possesses three oriented Weyl node pairs at $E_F = 60$ meV. The surface conductivity is normalized by $e^2/h$.

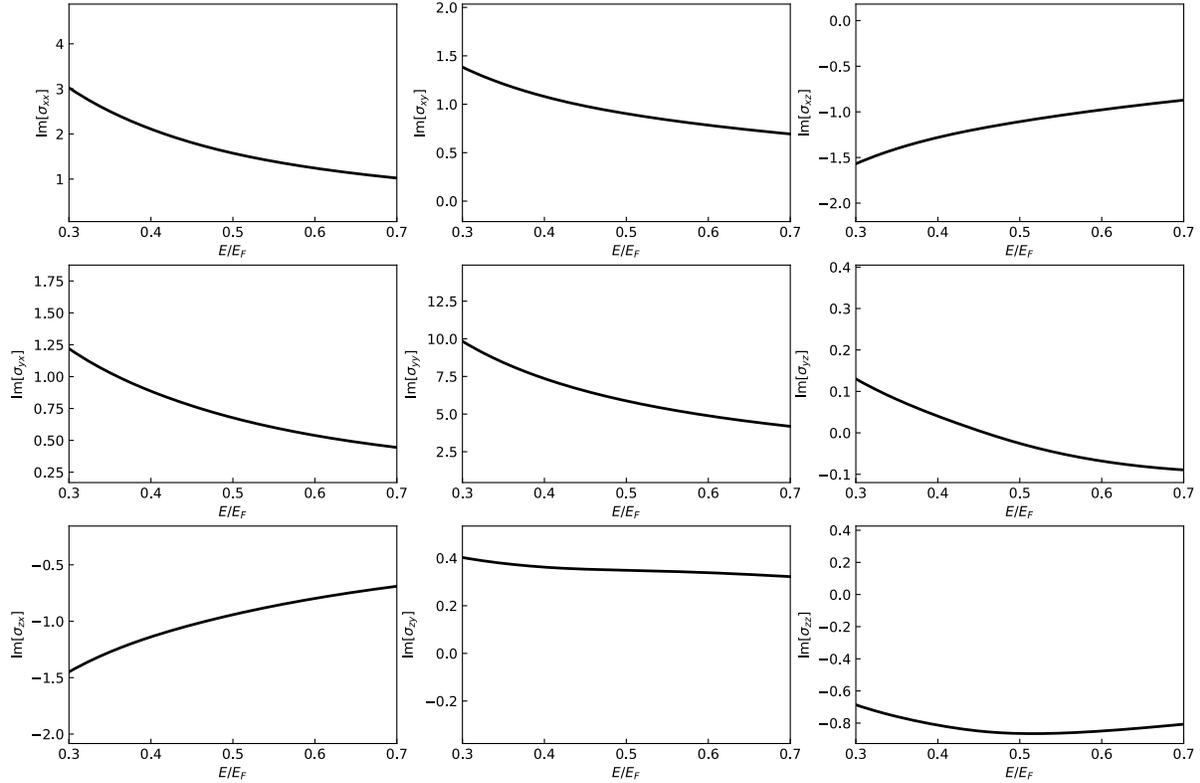

Figure S6.4: Imaginary parts of the elements of the surface conductivity tensor of a magnetic Weyl semimetal that possesses three oriented Weyl node pairs at $E_F = 60$ meV. The surface conductivity is normalized by $e^2/h$.

## References

[1] Q. Chen, A. R. Kutayiah, I. Oladyshkin, M. Tokman, and A. Belyanin, Phys. Rev. B **99**, 075137 (2019).

[2] R. Resta and D. Vanderbilt, Top. Appl. Phys. **105**, 31 (2007).

[3] Q. Wang, Y. Xu, R. Lou, Z. Liu, M. Li, Y. Huang, D. Shen, H. Weng, S. Wang, and H. Lei, Nat. Commun. **9**, 3681 (2018).